\pdfoutput=1
\newif\ifarxiv
\arxivtrue

\ifarxiv
  \documentclass[acmsmall,screen,acmthm,nonacm]{acmart}
\else
  \documentclass[acmsmall,screen,acmthm]{acmart}
\fi

\ifarxiv
    \setcopyright{cc}
\else
    \setcopyright{acmlicensed}
    \acmDOI{10.1145/3704840}
    \acmYear{2025}
    \acmJournal{PACMPL}
    \acmVolume{9}
    \acmNumber{POPL}
    \acmArticle{4}
    \acmMonth{1}
    \received{2024-07-10}
    \received[accepted]{2024-11-07}
\fi

\usepackage[utf8]{inputenc}

\usepackage[T1]{fontenc}
\usepackage[english]{babel}
\usepackage{csquotes}
\usepackage{booktabs,colortbl,multirow}
\usepackage{placeins}
\usepackage{subcaption}
\usepackage{wrapfig}
\usepackage{xspace}
\usepackage{mathtools}
\usepackage[obeyFinal]{todonotes}
\usepackage{afterpage}
\usepackage{stmaryrd}
\usepackage{enumitem}
\usepackage{nicefrac}
\usepackage{dashbox}
    \newcommand\dashedph[1][H]{\setlength{\fboxsep}{0pt}\setlength{\dashlength}{2.2pt}\setlength{\dashdash}{1.1pt} \dbox{\phantom{#1}}}

\usepackage{mathpartir}

\usepackage{acro}
    \newcommand{\newacro}[2]{\DeclareAcronym{#1}{short=#1,long=#2}}

\newacro{AD}{automatic differentiation}
\newacro{AST}{abstract syntax tree}
\newacro{CFG}{control-flow graph}
\newacro{CSE}{common subexpression elimination}
\newacro{DAG}{directed acyclic graph}
\newacro{DCE}{dead code elimination}
\newacro{DSL}{domain-specific language}
\newacro{GVN}{global value numbering}
\newacro{CPS}{continuation-passing style}
\newacro{LCSSA}{loop-closed SSA}
\newacro{JIT}{just-in-time}
\newacro{LoC}{lines of code}
\newacro{DOT}{dependent object types}
\newacro{NFA}{nondeterministic finite automaton}
\newacro{DFA}{deterministic finite automaton}
\newacro{ILP}{integer linear programming}

\DeclareAcronym{DTAL}{
    short = DTAL,
    long = Dependently Typed Assembly Language,
    cite = DBLP:conf/icfp/XiH01,
}

\DeclareAcronym{ArBB}{
    short = ArBB,
    long = Array Building Blocks,
    cite = DBLP:conf/cgo/NewburnSLMGTWDCWGLZ11,
}

\DeclareAcronym{CC}{
    short             = CC,
    long              = the Calculus of Constructions,
    cite              = DBLP:journals/iandc/CoquandH88,
}

\DeclareAcronym{CTRE}{
    short             = CTRE,
    long              = Compile Time Regular Expression,
    cite              = CTRE,
}

\DeclareAcronym{FFI}{
    short = FFI,
    long = foreign function interface,
    short-indefinite = an,
}

\DeclareAcronym{IR}{
    short             = IR,
    long              = intermediate representation,
    short-indefinite  = an,
}

\DeclareAcronym{PCRE2}{
    short             = PCRE2,
    long              = Perl-compatible Regular Expressions 2,
    cite              = pcre2,
}

\DeclareAcronym{PTS}{
    short             = PTS,
    short-plural-form = PTS,
    long              = pure type system,
    cite              = Bar:93,
}

\DeclareAcronym{RegEx}{
    short = RegEx,
    long = regular expression,
    short-plural = es,
}

\DeclareAcronym{SSA}{
    short             = SSA,
    long              = static single assignment,
    short-indefinite  = an,
}

\DeclareAcronym{SSI}{
    short             = SSI,
    long              = static single information,
    short-indefinite  = an,
}

\DeclareAcronym{SCCP}{
    short             = SCCP,
    long              = sparse conditional constant propagation,
    cite              = DBLP:conf/popl/WegmanZ85,
}

\DeclareAcronym{QTT}{
    short             = QTT,
    long              = Quantitative Type Theory,
    cite              = DBLP:conf/lics/Atkey18,
}

\usepackage{tikz}
    \usetikzlibrary{arrows,calc,positioning,shapes,matrix,decorations.pathreplacing,fpu,pgfplots.groupplots}

\usepackage{pgfplots}
    \pgfplotsset{compat=1.18}
    \usepackage{pgfplotstable}
    \usetikzlibrary{pgfplots.colorbrewer}

\definecolor{chartgreen}{named}{ACMGreen}
\definecolor{chartorange}{named}{ACMOrange}
\definecolor{chartpuruple}{named}{ACMPurple}
\definecolor{chartred}{named}{ACMRed}
\definecolor{chartyellow}{named}{ACMYellow}
\definecolor{chartlightblue}{named}{ACMLightBlue}
\definecolor{chartblue}{named}{ACMBlue}
\definecolor{chartldarkblue}{named}{ACMDarkBlue}

\usepackage{hhline} % used for the top-most horizontal line! in order to fix the top left corner cell empty
\usepackage{highlight}
\usepackage[pdftex,outline]{contour}
\usepackage{algorithm}

\usepackage{listings}
\usepackage{lstautogobble}
\usepackage{etoolbox}

\definecolor[named]{codegreen}    {named}{ACMGreen}
\definecolor[named]{codered}      {named}{ACMRed}
\definecolor[named]{codelightblue}{named}{ACMLightBlue}
\definecolor[named]{codedarkblue} {named}{ACMDarkBlue}

\expandafter\patchcmd\csname \string\lstinline\endcsname{%
    \leavevmode
    \bgroup
}{%
    \leavevmode
    \ifmmode\hbox\fi
    \bgroup
}{}{%
    \typeout{Patching of \string\lstinline\space failed!}%
}

    {\noindent\ignorespaces\begin{lstlisting}[#1,float,floatplacement=H]}{\end{lstlisting}\noindent\ignorespacesafterend}

\definecolor[named]{whitesmoke}   {rgb}{0.96,0.96,0.96}

\def\lst{\lstinline}

\makeatletter
\newenvironment{btHighlight}[1][]
{\begingroup\tikzset{bt@Highlight@par/.style={#1}}\begin{lrbox}{\@tempboxa}}
{\end{lrbox}\bt@HL@box[bt@Highlight@par]{\@tempboxa}\endgroup}

\newcommand\btHL[1][]{%
    \begin{btHighlight}[#1]\bgroup\aftergroup\bt@HL@endenv%
}
\def\bt@HL@endenv{%
    \end{btHighlight}%
    \egroup
}
\newcommand{\bt@HL@box}[2][]{%
    \tikz[#1]{%
        \pgfpathrectangle{\pgfpoint{1pt}{0pt}}{\pgfpoint{\wd #2}{\ht #2}}%
        \pgfusepath{use as bounding box}%
        \node[anchor=base west, fill=codelightblue,outer sep=0pt,inner xsep=1pt, inner ysep=0pt, rounded corners=3pt, minimum height=\ht\strutbox+1pt,#1]{\raisebox{1pt}{\strut}\strut\usebox{#2}};
    }%
}
\makeatother

\lstdefinelanguage{impala}{
    morecomment = [s]{/*}{*/},
    morecomment = [l]{//},
    sensitive = true,
    morekeywords = {i8,i16,i32,i64,u8,u16,u32,u64,f16,f32,f64,bool,int,float,double,extern,struct,as,match,true,false,type,with,let,mut,while,in,exit,return,break,continue,simd,if,else,for,do,fn,any,all,extract,shuffle,ballot,enum},
    moredelim=**[is][\btHL]{§}{§},
    morestring=[b]",
}

\lstdefinelanguage{mim}{
    alsoletter={.\%},
    literate=%
        {0_1}{{$0_1$}}1
        {0_2}{{$0_2$}}1
        {1_2}{{$1_2$}}1
        {0_3}{{$0_3$}}1
        {1_3}{{$1_3$}}1
        {2_3}{{$2_3$}}1
        {1_5}{{$1_5$}}1
        {->}{{$\shortrightarrow$}}1
        {<}{{\guilsinglleft}}1
        {>}{{\guilsinglright}}1
        {<<}{{\guillemetleft}}1
        {>>}{{\guillemetright}}1
        {BOT}{{$\color{ACMDarkBlue}\bot$}}1
        {TOP}{{$\color{ACMDarkBlue}\top$}}1
        {LAM}{{${\color{ACMDarkBlue}\lambda}$}}1
        {PI}{{${\color{ACMDarkBlue}\Pi}$}}1
        {PHI}{{$\phi$}}1,
    morecomment = [s]{/*}{*/},
    morecomment = [l]{//},
    sensitive = true,
    keywords = [1]{Type,axm,let,lam,lm,.Pi,.Sigma,con,cn,Cn,fun,fn,Fn,ins,Sort,Type,Idx,Bool,Cn,Nat,import,plugin,I2,I8,I16,I32,I64,f16,f32,f64,tt,ff,where,end},
    keywords = [2]{
        \%mem.M,\%mem.alloc,\%mem.Ptr,\%mem.load,\%mem.store,\%mem.lea,\%mem.Ptr,\%mem.Ptr0,\%mem.free,\%mem.slot,%
        \%mem.ssa_pass,
        \%autodiff.ad,
        \%autodiff.AD,
        \%core.ncmp,\%core.ncmp.e,\%core.ncmp.l,
        \%core.nat,\%core.nat.add,\%core.nat.sub,\%core.nat.mul,
        \%core.bit1,
        \%core.bit2,
        \%core.wrap,\%core.wrap.add,\%core.wrap.sub,\%core.wrap.mul,\%core.div.sdiv,\%core.div.udiv,\%core.bit2.and_,
        \%core.shr,
        \%core.idx,
        \%core.pe.known,
        \%core.bitcast,\%core.conv,\%core.conv.u,
        \%core.pe.know,
        \%core.minus,
        \%core.icmp,\%core.icmp.ule,\%core.icmp.uge,\%core.icmp.e,
        \%compile.Pass,
        \%compile.Phase,
        \%compile.pipe,
        \%compile.Pipeline,
        \%compile.phase_list,
        \%compile.combined_phase,
        \%compile.CombinedPhase,
        \%compile.pass_list,
        \%compile.pass_phase,
        \%compile.PassList,
        \%direct.cps2ds,
        \%math.F,\%math.f64,\%math.F64,
        \%math.arith,\%math.arith.add,\%math.arith.sub,\%math.arith.mul,\%math.arith.div,\%math.arith.mod,
        \%math.tri,
        \%math.cmp,
        \%math.minus,
        \%tensor.Mat,
        \%tensor.zip,
        \%tensor.reduce,
        \%regex.lit,\%regex.range,\%regex.math_range,
        \%regex.cls.w,\%regex.cls.W,
        \%regex.not_,\%regex.any,\%regex.disj,\%regex.conj,\%regex.not,\%regex.cls.d,\%regex.cls.D,
        \%regex.quant,\%regex.quant.plus,\%regex.quant.star,\%regex.quant.optional,
        \%regex.lower_regex,\%regex.match_range,
        \%tensor.zip,\%tensor.reduce,\%tensor.map,\%affine.For,
        \%refly.reify,\%refly.reflect,\%refly.refine,\%refly.Code},
    moredelim=**[is][\btHL]{§}{§},
}

\lstdefinelanguage{Rust}{%
  sensitive%
, morecomment=[l]{//}%
, morecomment=[s]{/*}{*/}%
, moredelim=[s][{\itshape\color[rgb]{0,0,0.75}}]{\#[}{]}%
, morestring=[b]{"}%
, alsodigit={}%
, alsoother={}%
, alsoletter={!}%
, morekeywords={break, continue, else, for, if, in, loop, match, return, while}  % control flow keywords
, morekeywords={as, const, let, move, mut, ref, static}  % in the context of variables
, morekeywords={dyn, enum, fn, impl, Self, self, struct, trait, type, union, use, where}  % in the context of declarations
, morekeywords={crate, extern, mod, pub, super}  % in the context of modularisation
, morekeywords={unsafe}  % markers
, morekeywords={abstract, alignof, become, box, do, final, macro, offsetof, override, priv, proc, pure, sizeof, typeof, unsized, virtual, yield}  % reserved identifiers
, morekeywords=[2]{Add, AddAssign, Any, AsciiExt, AsInner, AsInnerMut, AsMut, AsRawFd, AsRawHandle, AsRawSocket, AsRef, Binary, BitAnd, BitAndAssign, Bitor, BitOr, BitOrAssign, BitXor, BitXorAssign, Borrow, BorrowMut, Boxed, BoxPlace, BufRead, BuildHasher, CastInto, CharExt, Clone, CoerceUnsized, CommandExt, Copy, Debug, DecodableFloat, Default, Deref, DerefMut, DirBuilderExt, DirEntryExt, Display, Div, DivAssign, DoubleEndedIterator, DoubleEndedSearcher, Drop, EnvKey, Eq, Error, ExactSizeIterator, ExitStatusExt, Extend, FileExt, FileTypeExt, Float, Fn, FnBox, FnMut, FnOnce, Freeze, From, FromInner, FromIterator, FromRawFd, FromRawHandle, FromRawSocket, FromStr, FullOps, FusedIterator, Generator, Hash, Hasher, Index, IndexMut, InPlace, Int, Into, IntoCow, IntoInner, IntoIterator, IntoRawFd, IntoRawHandle, IntoRawSocket, IsMinusOne, IsZero, Iterator, JoinHandleExt, LargeInt, LowerExp, LowerHex, MetadataExt, Mul, MulAssign, Neg, Not, Octal, OpenOptionsExt, Ord, OsStrExt, OsStringExt, Packet, PartialEq, PartialOrd, Pattern, PermissionsExt, Place, Placer, Pointer, Product, Put, RangeArgument, RawFloat, Read, Rem, RemAssign, Seek, Shl, ShlAssign, Shr, ShrAssign, Sized, SliceConcatExt, SliceExt, SliceIndex, Stats, Step, StrExt, Sub, SubAssign, Sum, Sync, TDynBenchFn, Terminal, Termination, ToOwned, ToSocketAddrs, ToString, Try, TryFrom, TryInto, UnicodeStr, Unsize, UpperExp, UpperHex, WideInt, Write}
, morekeywords=[2]{Send}  % additional traits
, morekeywords=[3]{bool, char, f32, f64, i8, i16, i32, i64, isize, str, u8, u16, u32, u64, unit, usize, i128, u128}  % primitive types
, morekeywords=[4]{Err, false, None, Ok, Some, true}  % prelude value constructors
, morekeywords=[3]{AccessError, Adddf3, AddI128, AddoI128, AddoU128, ADDRESS, ADDRESS64, addrinfo, ADDRINFOA, AddrParseError, Addsf3, AddU128, advice, aiocb, Alignment, AllocErr, AnonPipe, Answer, Arc, Args, ArgsInnerDebug, ArgsOs, Argument, Arguments, ArgumentV1, Ashldi3, Ashlti3, Ashrdi3, Ashrti3, AssertParamIsClone, AssertParamIsCopy, AssertParamIsEq, AssertUnwindSafe, AtomicBool, AtomicPtr, Attr, auxtype, auxv, BackPlace, BacktraceContext, Barrier, BarrierWaitResult, Bencher, BenchMode, BenchSamples, BinaryHeap, BinaryHeapPlace, blkcnt, blkcnt64, blksize, BOOL, boolean, BOOLEAN, BoolTrie, BorrowError, BorrowMutError, Bound, Box, bpf, BTreeMap, BTreeSet, Bucket, BucketState, Buf, BufReader, BufWriter, Builder, BuildHasherDefault, BY, BYTE, Bytes, CannotReallocInPlace, cc, Cell, Chain, CHAR, CharIndices, CharPredicateSearcher, Chars, CharSearcher, CharsError, CharSliceSearcher, CharTryFromError, Child, ChildPipes, ChildStderr, ChildStdin, ChildStdio, ChildStdout, Chunks, ChunksMut, ciovec, clock, clockid, Cloned, cmsgcred, cmsghdr, CodePoint, Color, ColorConfig, Command, CommandEnv, Component, Components, CONDITION, condvar, Condvar, CONSOLE, CONTEXT, Count, Cow, cpu, CRITICAL, CStr, CString, CStringArray, Cursor, Cycle, CycleIter, daddr, DebugList, DebugMap, DebugSet, DebugStruct, DebugTuple, Decimal, Decoded, DecodeUtf16, DecodeUtf16Error, DecodeUtf8, DefaultEnvKey, DefaultHasher, dev, device, Difference, Digit32, DIR, DirBuilder, dircookie, dirent, dirent64, DirEntry, Discriminant, DISPATCHER, Display, Divdf3, Divdi3, Divmoddi4, Divmodsi4, Divsf3, Divsi3, Divti3, dl, Dl, Dlmalloc, Dns, DnsAnswer, DnsQuery, dqblk, Drain, DrainFilter, Dtor, Duration, DwarfReader, DWORD, DWORDLONG, DynamicLibrary, Edge, EHAction, EHContext, Elf32, Elf64, Empty, EmptyBucket, EncodeUtf16, EncodeWide, Entry, EntryPlace, Enumerate, Env, epoll, errno, Error, ErrorKind, EscapeDebug, EscapeDefault, EscapeUnicode, event, Event, eventrwflags, eventtype, ExactChunks, ExactChunksMut, EXCEPTION, Excess, ExchangeHeapSingleton, exit, exitcode, ExitStatus, Failure, fd, fdflags, fdsflags, fdstat, ff, fflags, File, FILE, FileAttr, filedelta, FileDesc, FilePermissions, filesize, filestat, FILETIME, filetype, FileType, Filter, FilterMap, Fixdfdi, Fixdfsi, Fixdfti, Fixsfdi, Fixsfsi, Fixsfti, Fixunsdfdi, Fixunsdfsi, Fixunsdfti, Fixunssfdi, Fixunssfsi, Fixunssfti, Flag, FlatMap, Floatdidf, FLOATING, Floatsidf, Floatsisf, Floattidf, Floattisf, Floatundidf, Floatunsidf, Floatunsisf, Floatuntidf, Floatuntisf, flock, ForceResult, FormatSpec, Formatted, Formatter, Fp, FpCategory, fpos, fpos64, fpreg, fpregset, FPUControlWord, Frame, FromBytesWithNulError, FromUtf16Error, FromUtf8Error, FrontPlace, fsblkcnt, fsfilcnt, fsflags, fsid, fstore, fsword, FullBucket, FullBucketMut, FullDecoded, Fuse, GapThenFull, GeneratorState, gid, glob, glob64, GlobalDlmalloc, greg, group, GROUP, Guard, GUID, Handle, HANDLE, Handler, HashMap, HashSet, Heap, HINSTANCE, HMODULE, hostent, HRESULT, id, idtype, if, ifaddrs, IMAGEHLP, Immut, in, in6, Incoming, Infallible, Initializer, ino, ino64, inode, input, InsertResult, Inspect, Instant, int16, int32, int64, int8, integer, IntermediateBox, Internal, Intersection, intmax, IntoInnerError, IntoIter, IntoStringError, intptr, InvalidSequence, iovec, ip, IpAddr, ipc, Ipv4Addr, ipv6, Ipv6Addr, Ipv6MulticastScope, Iter, IterMut, itimerspec, itimerval, jail, JoinHandle, JoinPathsError, KDHELP64, kevent, kevent64, key, Key, Keys, KV, l4, LARGE, lastlog, launchpad, Layout, Lazy, lconv, Leaf, LeafOrInternal, Lines, LinesAny, LineWriter, linger, linkcount, LinkedList, load, locale, LocalKey, LocalKeyState, Location, lock, LockResult, loff, LONG, lookup, lookupflags, LookupHost, LPBOOL, LPBY, LPBYTE, LPCSTR, LPCVOID, LPCWSTR, LPDWORD, LPFILETIME, LPHANDLE, LPOVERLAPPED, LPPROCESS, LPPROGRESS, LPSECURITY, LPSTARTUPINFO, LPSTR, LPVOID, LPWCH, LPWIN32, LPWSADATA, LPWSAPROTOCOL, LPWSTR, Lshrdi3, Lshrti3, lwpid, M128A, mach, major, Map, mcontext, Metadata, Metric, MetricMap, mflags, minor, mmsghdr, Moddi3, mode, Modsi3, Modti3, MonitorMsg, MOUNT, mprot, mq, mqd, msflags, msghdr, msginfo, msglen, msgqnum, msqid, Muldf3, Mulodi4, Mulosi4, Muloti4, Mulsf3, Multi3, Mut, Mutex, MutexGuard, MyCollection, n16, NamePadding, NativeLibBoilerplate, nfds, nl, nlink, NodeRef, NoneError, NonNull, NonZero, nthreads, NulError, OccupiedEntry, off, off64, oflags, Once, OnceState, OpenOptions, Option, Options, OptRes, Ordering, OsStr, OsString, Output, OVERLAPPED, Owned, Packet, PanicInfo, Param, ParseBoolError, ParseCharError, ParseError, ParseFloatError, ParseIntError, ParseResult, Part, passwd, Path, PathBuf, PCONDITION, PCONSOLE, Peekable, PeekMut, Permissions, PhantomData, pid, Pipes, PlaceBack, PlaceFront, PLARGE, PoisonError, pollfd, PopResult, port, Position, Powidf2, Powisf2, Prefix, PrefixComponent, PrintFormat, proc, Process, PROCESS, processentry, protoent, PSRWLOCK, pthread, ptr, ptrdiff, PVECTORED, Queue, radvisory, RandomState, Range, RangeFrom, RangeFull, RangeInclusive, RangeMut, RangeTo, RangeToInclusive, RawBucket, RawFd, RawHandle, RawPthread, RawSocket, RawTable, RawVec, Rc, ReadDir, Receiver, recv, RecvError, RecvTimeoutError, ReentrantMutex, ReentrantMutexGuard, Ref, RefCell, RefMut, REPARSE, Repeat, Result, Rev, Reverse, riflags, rights, rlim, rlim64, rlimit, rlimit64, roflags, Root, RSplit, RSplitMut, RSplitN, RSplitNMut, RUNTIME, rusage, RwLock, RWLock, RwLockReadGuard, RwLockWriteGuard, sa, SafeHash, Scan, sched, scope, sdflags, SearchResult, SearchStep, SECURITY, SeekFrom, segment, Select, SelectionResult, sem, sembuf, send, Sender, SendError, servent, sf, Shared, shmatt, shmid, ShortReader, ShouldPanic, Shutdown, siflags, sigaction, SigAction, sigevent, sighandler, siginfo, Sign, signal, signalfd, SignalToken, sigset, sigval, Sink, SipHasher, SipHasher13, SipHasher24, size, SIZE, Skip, SkipWhile, Slice, SmallBoolTrie, sockaddr, SOCKADDR, sockcred, Socket, SOCKET, SocketAddr, SocketAddrV4, SocketAddrV6, socklen, speed, Splice, Split, SplitMut, SplitN, SplitNMut, SplitPaths, SplitWhitespace, spwd, SRWLOCK, ssize, stack, STACKFRAME64, StartResult, STARTUPINFO, stat, Stat, stat64, statfs, statfs64, StaticKey, statvfs, StatVfs, statvfs64, Stderr, StderrLock, StderrTerminal, Stdin, StdinLock, Stdio, StdioPipes, Stdout, StdoutLock, StdoutTerminal, StepBy, String, StripPrefixError, StrSearcher, subclockflags, Subdf3, SubI128, SuboI128, SuboU128, subrwflags, subscription, Subsf3, SubU128, Summary, suseconds, SYMBOL, SYMBOLIC, SymmetricDifference, SyncSender, sysinfo, System, SystemTime, SystemTimeError, Take, TakeWhile, tcb, tcflag, TcpListener, TcpStream, TempDir, TermInfo, TerminfoTerminal, termios, termios2, TestDesc, TestDescAndFn, TestEvent, TestFn, TestName, TestOpts, TestResult, Thread, threadattr, threadentry, ThreadId, tid, time, time64, timespec, TimeSpec, timestamp, timeval, timeval32, timezone, tm, tms, ToLowercase, ToUppercase, TraitObject, TryFromIntError, TryFromSliceError, TryIter, TryLockError, TryLockResult, TryRecvError, TrySendError, TypeId, U64x2, ucontext, ucred, Udivdi3, Udivmoddi4, Udivmodsi4, Udivmodti4, Udivsi3, Udivti3, UdpSocket, uid, UINT, uint16, uint32, uint64, uint8, uintmax, uintptr, ulflags, ULONG, ULONGLONG, Umoddi3, Umodsi3, Umodti3, UnicodeVersion, Union, Unique, UnixDatagram, UnixListener, UnixStream, Unpacked, UnsafeCell, UNWIND, UpgradeResult, useconds, user, userdata, USHORT, Utf16Encoder, Utf8Error, Utf8Lossy, Utf8LossyChunk, Utf8LossyChunksIter, utimbuf, utmp, utmpx, utsname, uuid, VacantEntry, Values, ValuesMut, VarError, Variables, Vars, VarsOs, Vec, VecDeque, vm, Void, WaitTimeoutResult, WaitToken, wchar, WCHAR, Weak, whence, WIN32, WinConsole, Windows, WindowsEnvKey, winsize, WORD, Wrapping, wrlen, WSADATA, WSAPROTOCOL, WSAPROTOCOLCHAIN, Wtf8, Wtf8Buf, Wtf8CodePoints, xsw, xucred, Zip, zx}
, morekeywords=[5]{assert!, assert_eq!, assert_ne!, cfg!, column!, compile_error!, concat!, concat_idents!, debug_assert!, debug_assert_eq!, debug_assert_ne!, env!, eprint!, eprintln!, file!, format!, format_args!, include!, include_bytes!, include_str!, line!, module_path!, option_env!, panic!, print!, println!, select!, stringify!, thread_local!, try!, unimplemented!, unreachable!, vec!, write!, writeln!}  % prelude macros
}%

\def\to{\rightarrow}

\def\hlgamma{\hl{$\Gamma$}}

\def\hlstar{\hl{*}}
\def\rto{\leftarrow}

\def\vtr{\ensuremath{\triangleright}}

\def\hole{\ensuremath{\boldsymbol{\cdot}}}
\newcommand{\arr}[1]{\text{\guillemetleft}#1\text{\guillemetright}}
\newcommand{\pack}[1]{\text{\guilsinglleft}#1\text{\guilsinglright}}
\newcommand{\key}[1]{{\color{ACMDarkBlue}\textbf{\texttt{#1}}}}
\newcommand{\hl}[1]{{\texttt{\color{ACMBlue}#1}}}
\newcommand{\mhl}[1]{\mbox{\hl{#1}}}
\newcommand{\ul}[1]{\underline{#1}}
\newcommand{\hul}[1]{\hl{\ul{#1}}}
\newcommand{\type}[2]{\hlgamma \vdash \hl{#1} : \hl{#2}}

\newcommand{\atype}[2]{\hlgamma \vdash \hl{#1} \rto \hl{#2}}
\newcommand{\step}[2]{\hl{#1} \to \hl{#2}}
\newcommand{\irule}[3]{\mbox{\hypertarget{rule:#1}{\textsc{#1}}\inferrule{#2}{#3}}}
\newcommand{\irulex}[4]{\mbox{\hypertarget{rule:#1}{\textsc{#2}}\inferrule{#3}{#4}}}
\newcommand{\iref}[1]{\hyperlink{rule:#1}{\textsc{#1}}}
\newcommand{\irefrule}[1]{\hyperlink{rule:#1}{Rule~\textsc{#1}}}
\newcommand{\irefx}[2]{\hyperlink{rule:#1}{\textsc{#2}}}
\newcommand{\irefrulex}[2]{\hyperlink{rule:#1}{Rule~\textsc{#2}}}
\newcommand{\subst}[2]{[\hl{#1}/\hl{#2}]}
\newcommand{\FV}[1]{\ensuremath{\mathit{FV}\{\hl{#1}\}}}
\newcommand\mydots{\hbox to 1em{.\hss.\hss.}}
\def\Mid{\ \,\mid\ \,}

\newcommand{\toolname}[1]{\textsc{#1}\@\xspace}
\newcommand{\adbench}{\toolname{ADBench}}
\newcommand{\blas}{\toolname{BLAS}}
\newcommand{\clang}{\toolname{Clang}}
\newcommand{\enzyme}{\toolname{Enzyme}}

\newcommand{\impala}{\toolname{Impala}}
\newcommand{\llvm}{\toolname{LLVM}}
\newcommand{\mlir}{\toolname{MLIR}}

\newcommand{\pytorch}{\toolname{PyTorch}}
\newcommand{\qqtext}[1]{\qquad\text{#1}\qquad}
\newcommand{\qtext}[1]{\quad\text{#1}\quad}

\newcommand{\mimir}{\toolname{MimIR}}
\newcommand{\mim}{\toolname{Mim}}
\newcommand{\thorin}{\toolname{Thorin}}
\newcommand{\thornado}{\toolname{Thornado}}
\newcommand{\torchscript}{\toolname{TorchScript}}
\newcommand{\circt}{\toolname{CIRCT}}

\makeatletter
\DeclareFontFamily{OMX}{MnSymbolE}{}
\DeclareSymbolFont{MnLargeSymbols}{OMX}{MnSymbolE}{m}{n}
\SetSymbolFont{MnLargeSymbols}{bold}{OMX}{MnSymbolE}{b}{n}
\DeclareFontShape{OMX}{MnSymbolE}{m}{n}{
    <-6>  MnSymbolE5
   <6-7>  MnSymbolE6
   <7-8>  MnSymbolE7
   <8-9>  MnSymbolE8
   <9-10> MnSymbolE9
  <10-12> MnSymbolE10
  <12->   MnSymbolE12
}{}
\DeclareFontShape{OMX}{MnSymbolE}{b}{n}{
    <-6>  MnSymbolE-Bold5
   <6-7>  MnSymbolE-Bold6
   <7-8>  MnSymbolE-Bold7
   <8-9>  MnSymbolE-Bold8
   <9-10> MnSymbolE-Bold9
  <10-12> MnSymbolE-Bold10
  <12->   MnSymbolE-Bold12
}{}

\let\llangle\@undefined
\let\rrangle\@undefined
\DeclareMathDelimiter{\llangle}{\mathopen}%
                     {MnLargeSymbols}{'164}{MnLargeSymbols}{'164}
\DeclareMathDelimiter{\rrangle}{\mathclose}%
                     {MnLargeSymbols}{'171}{MnLargeSymbols}{'171}
\makeatother
\begin{document}

\title{MimIR: An Extensible and Type-Safe Intermediate Representation for the DSL Age}

\author{Roland Leißa}
\orcid{0000-0002-2444-6782}
\affiliation{%
  \institution{University of Mannheim}
  \city{Mannheim}
  \country{Germany}
}
\email{leissa@uni-mannheim.de}

\author{Marcel Ullrich}
\orcid{0009-0006-0127-9623}
\affiliation{%
  \institution{Saarland University}
  \city{Saarbrücken}
  \country{Germany}
}
\email{ullrich@cs.uni-saarland.de}

\author{Joachim Meyer}
\orcid{0000-0003-2656-9863}
\affiliation{%
  \institution{Saarland University}
  \city{Saarbrücken}
  \country{Germany}
}
\email{jmeyer@cs.uni-saarland.de}

\author{Sebastian Hack}
\orcid{0000-0002-3387-2134}
\affiliation{%
  \institution{Saarland University}
  \city{Saarbrücken}
  \country{Germany}
}
\email{hack@cs.uni-saarland.de}

%% Abstract
%% Note: \begin{abstract}...\end{abstract} environment must come
%% before \maketitle command
% vim:spell:spelllang=en
\begin{abstract}
    Traditional compilers, designed for optimizing low-level code, fall short when dealing with modern, computation-heavy applications like image processing, machine learning, or numerical simulations.
    Optimizations should understand the primitive operations of the specific application domain and thus happen on that level.

    Domain-specific languages (DSLs) fulfill these requirements.
    However, DSL compilers reinvent the wheel over and over again as standard optimizations, code generators, and general infrastructure \& boilerplate code must be reimplemented for each DSL compiler.

    This paper presents \mimir, an extensible, higher-order intermediate representation.
    At its core, \mimir is a pure type system and, hence, a form of a typed lambda calculus.
    Developers can declare the signatures of new (domain-specific) operations, called \emph{axioms}.
    An axiom can be the declaration of a function, a type constructor, or any other entity with a possibly polymorphic, polytypic, and/or dependent type.
    This way, developers can extend \mimir at any low or high level and bundle them in a \emph{plugin}.
    Plugins extend the compiler and take care of optimizing and lowering the plugins' axioms.

    We show the expressiveness and effectiveness of \mimir in three case studies:
    Low-level plugins that operate at the same level of abstraction as \llvm, a regular-expression matching plugin, and plugins for linear algebra and automatic differentiation.
    We show that in all three studies, \mimir produces code that has state-of-the-art performance.
\end{abstract}

%% 2012 ACM Computing Classification System (CSS) concepts
%% Generate at 'http://dl.acm.org/ccs/ccs.cfm'.



\maketitle

% vim:spell:spelllang=en
\section{Introduction}
\label{sec:intro}

Dennard scaling enabled the continuous growth of single-thread performance of legacy code for decades.
After its decline in the early 2000s, \acp{DSL} have received a lot of attention in the effort to harvest as much performance as possible in a productive way.
\acp{DSL} offer specialized abstractions tailored to specific problem domains enhancing the programmer's productivity on one side and enabling the generation of high-performance code on the other side.
To generate actual code, \ac{DSL} compilers typically resort to existing compiler frameworks such as \llvm~\cite{DBLP:conf/cgo/LattnerA04}.
However, there is a significant gap in the abstractions \acp{DSL} provide and the low-level nature of these compiler frameworks which often leads to the situation that \ac{DSL} compilers have an additional, more high-level \ac{IR} that bridges the gap between the \ac{DSL} and the back-end compiler.

This more high-level \ac{IR} is often designed and implemented for each new \ac{DSL} from scratch.
Current research in high-level \acp{IR}, most notably \mlir~\cite{DBLP:conf/cgo/LattnerAB0DPRSV21}, seeks to provide a least common denominator for the basis of such an \ac{IR} to facilitate the reuse of core data structures and algorithms across different \acp{DSL} but not much more.
While \mlir provides extensibility to host domain-specific \emph{dialects}, it is still low-level in several aspects.
First, it relies on \acp{CFG}, thereby relying on multiple concepts to represent control flow including functions, basic blocks, instructions, and so-called regions to delineate code for high-level transformations.
Second, \mlir does not provide a type system that is expressive enough to host the type systems of \acp{DSL}.
It relies on the dialects to implement their own type systems in C++.
This causes more implementation effort for the \ac{DSL} implementer and an unclear notion of well-typedness when multiple dialects interact.
Finally, because \mlir is designed with \enquote{little builtin, everything customizable}~\cite[p. 3]{DBLP:conf/cgo/LattnerAB0DPRSV21}, it is \emph{by design} impossible to give a formal account of its static and dynamic semantics that would encompass all \acp{DSL}.
% The type system therefore can give fewer guarantees regarding the semantics of and between individual \acp{DSL}.

In this paper, we want to go one step further and present \mimir, a higher-order \ac{IR} that provides a type system that sets out to be expressive enough to host a wide range of \acp{DSL}.
Consider the following \ac{DSL} interface for a \emph{tensor} plugin in \mimir's \emph{surface language} \mim---the textual interface of \mimir:
\begin{lstlisting}
    axm %tensor.Mat: {n m: Nat, T: *} -> *; // n $\color{ACMGreen}\times$ m matrix with element type T
    axm %tensor.zip: {n m: Nat, T: *} [f: [T, T] -> T] [a b: %tensor.Mat (n, m, T)]
                                                          -> %tensor.Mat (n, m, T);
\end{lstlisting}
The so-called \emph{axiom} \hl{\%tensor.Mat} is a type constructor that expects two dimensions~\hl{n} and \hl{m} and an element type \hl{T} to construct a matrix.
The \hl{\%tensor.zip} axiom is polymorphic in the arguments' dimensions and element type (whose values will be inferred at a call-site as the curly braces mark this parameter group as \emph{implicit}) and computes a new matrix by applying \hl{f} in pairs to all elements of both inputs.
Axioms are declarations of entities that do not have an implementation in \mimir per se.
Instead, \ac{DSL} designers define a \emph{plugin} that consists of an interface (see above) and a C++ implementation that provides domain-specific program transformations and code generators
to lower the axioms into the language of lower-level \mimir plugins or to generate target code directly.

\acp{IR} like \llvm are not flexible enough to express types like \hl{\%tensor.Mat} at all.
In \mlir, the designers of plugins (there called dialects) need to manually implement the type system in C++ for each dialect operation.
But even this has many severe limitations as \mlir does neither directly support higher-order functions nor polymorphism.
%For example, \texttt{vector.reduction}\footnote{see \url{https://mlir.llvm.org/docs/Dialects/Vector/\#vectorreduction-vectorreductionop}} from the \mlir \emph{vector} dialect hard-codes a set of predefined reduction operations.
For example, \href{https://mlir.llvm.org/docs/Dialects/Vector/\#vectorreduction-vectorreductionop}{\texttt{vector.reduction}} from the \mlir \emph{vector} dialect hard-codes a set of predefined reduction operations.
In \mimir we can define a \hl{reduce} function that works on any array and appropriate function.
For example, here we create a matrix addition with \hl{\%tensor.zip elem\_add} that we use to reduce an \hl{array\_of\_matrices} to a single \mbox{\hl{res\_mat}rix}:
\begin{lstlisting}
    let res_mat = reduce (%tensor.zip elem_add) (init, array_of_matrices);
\end{lstlisting}

Note that \hl{\%tensor.Mat}'s type is an ordinary function type and, yet, it is a type operator.
Furthermore, note that the types of \hl{\%tensor.zip}'s arguments \hl{a} and \hl{b} depend on the type variable~\hl{T} (polymorphism) and the term variables~\hl{n} and \hl{m} which makes \hl{\%tensor.Mat (n, m, T)} a \emph{dependent type}.
This is possible because at its core \mimir is a minimal, typed $\lambda$-calculus that is based upon the $\lambda$-cube~\cite{DBLP:journals/jfp/Barendregt91}, \ac{CC}, and \ac{PTS}.
%Moreover, \mimir borrows ideas from the zip calculus~\cite{DBLP:conf/mpc/Tullsen00} to allow for abstraction over tuple/array length which in turn enables variadic functions \& genericity over array rank.
This makes \mimir not only very expressive but also simplifies the compiler design in many ways because \mimir only knows a single syntactic category: expressions.
However, \mim supports syntactic sugar that \mim translates into its minimal, graph-based representation \mimir.

\subsection{Contributions}
In summary, this paper makes the following contributions:

\begin{itemize}[left=0pt .. \parindent]
    \item We introduce the \acl{IR} \mimir that allows, through its plugin architecture, to modularly extend the \mimir compiler with custom, domain-specific operations, types, and type operators, as well as domain-specific code transformations---in particular so-called \emph{normalizations} that are eagerly applied by \mimir (\autoref{sec:overview}).

    \item We formally introduce \mimir's semantics and its sound type system which allows for a type-safe composition of plugins.
        \mimir's expressive type system is rooted in \acl{CC} and, hence, features dependent-types.
        These are mainly used in proof assistants and associated with writing complex proof terms.
        \mimir, however, features full recursive functions and dependent types without the hassle of writing complex proof expressions.
        This is possible because \mimir's type checker tightly interacts with normalizations and a partial evaluator which allows for a novel way to type-check dependent types (\autoref{sec:sema}).

    \item Code transformations are implemented in C++ by manipulating \mimir's program graph.
        This is a \enquote{sea of nodes}-style~\cite{DBLP:conf/irep/ClickP95} \ac{IR} for a higher-order, \ac{PTS}-style language.
        As \mimir only knows expressions, the program graph is also very simple:
        Each node represents an expression and has outgoing edges to each of its operands and the type it inhabits---which is again an expression.
        Besides an internal hash set that hash-conses all nodes, \mimir does \emph{not} need any other auxiliary data structures such as instruction lists, basic blocks, \acp{CFG}, or special regions~(\autoref{sec:graph}).

    \item \mimir performs type checking, inference, and normalization on-the-fly \emph{during program construction}.
        This way, a plugin/compiler developer can resort to type inference, normalization, and (partial) evaluation upon creating an expression---regardless of whether they use \mim or \mimir's C++ interface.
        \mimir will immediately report any typing errors (\autoref{sec:impl}).

    \item \autoref{sec:eval} presents case studies that show \mimir's versatility and ability to host \acp{DSL} and generate high-performance code.
        These include
        a set of low-level, \llvm-like plugins, that we show to perform just as well as if directly using \llvm,
        a plugin for matching \acp{RegEx} which outperforms other popular \ac{RegEx} engines,
        a plugin for tensor computations, and
        a plugin for automatic differentiation with state-of-the-art performance at a tenth of the complexity.
\end{itemize}

% vim:spell:spelllang=en
\section{Overview}
\label{sec:overview}

The core of \mimir is a minimal, typed $\lambda$-calculus with strong semantics based on the $\lambda$-cube, \ac{PTS}, and \ac{CC}. % notably CC and PTS, so that stuff can be plugged together nicely..
We describe the core calculus' syntax and semantics in more detail in \autoref{sec:sema}.
What makes \mimir an attractive target for \acp{DSL} is its extensibility through plugins.
These plugins can define intrinsic operations (in \mimir called \enquote{axioms}) on various abstraction levels.
Furthermore, plugins can provide transformations and lowering passes to perform (domain-specific) optimizations on each abstraction level and to convert between the levels.
There are several ways how \iac{DSL} or a general-purpose language can target \mimir.
The most important ones are:
an embedded \ac{DSL} or \iac{DSL} compiler may use \mimir's API to construct the \ac{IR}.
Alternatively, a compiler can communicate with \mimir textually through its surface language \mim.
\mimir itself comes with a backend that emits textual \llvm IR.
This IR can be further processed by \llvm tools to obtain an executable.
Thereby, \mimir currently targets any CPU supported by \llvm.
Since plugins are essential to \mimir's design, we will use the example of a plugin for matching \acp{RegEx} to present \mimir's components.

\subsection{Plugin Architecture}

A plugin consists of two parts:
A \verb|*.mim| file that contains declarations in \mim of what the plugin exports and code transformations implemented in C++. % (see \autoref{fig:overview}).
The \texttt{\%regex} plugin declares its operations in \verb|regex.mim| (\autoref{lst:regex-plugin}).
% Keywords are prefixed with a dot to avoid name clashes with ordinary identifiers.
All names prefixed with \texttt{\%regex} are called \emph{annexes}.
Before building the plugin's C++ sources, \mimir \emph{bootstraps} the plugin by parsing \verb|regex.mim| and generating a C++ header file that declares all annexes as C++ \lst[language=C++]|enum|s.
This process does \emph{not} involve a \enquote{compilation} of the annexes to C++ in any way.
The purpose of this is that the C++ part of the plugin can reference an annex via a C++ name instead of a string (see below).
Then, the build system compiles the plugin's sources with the help of the generated header into a shared object \verb|libmim_regex.so|.
Now, other \mimir code can use the plugin (see \autoref{lst:ptrn}):
The \hl{\key{plugin} regex} directive instructs \mimir to parse \verb|regex.mim|.
In doing so, \mimir will know the names of all \texttt{\%regex} annexes and their types.
In addition, \mimir will dynamically load \verb|libmim_regex.so| to incorporate the plugin's code transformations into the \mimir compiler.
Now that the plugin has been loaded, the \lstinline|let|-expression binds the variable \hl{pattern} to a \mimir expression that represents the \ac{RegEx} \verb|^\w+\.[a-z]+$| (which matches simple top-level domains).

\begin{lstlisting}[
    label=lst:regex-plugin,
    caption=RegEx plugin declaration file \texttt{regex.mim},
    float=tp,
    firstline=1,
    numbers=right,
    numbersep=-2.5ex,
    abovecaptionskip=-.5ex,
    belowcaptionskip=-2ex]
    let Char            = I8;                                            $\label{line:regex-Char}$
    lam Str (n: Nat): * = %mem.Ptr <<n; Char>>;                           $\label{line:regex-Str}$
    lam Res (n: Nat): * = [%mem.M, Bool, Idx n];
    let RegEx           = {n: Nat} [%mem.M, Str n, Idx n] -> Res n;

    axm %regex.any:                               RegEx;
    axm %regex.conj:      {i: Nat} [<<i; RegEx>>] -> RegEx, normalize_conj,  2;
    axm %regex.disj:      {i: Nat} [<<i; RegEx>>] -> RegEx, normalize_disj,  2;
    axm %regex.range:                 <<2; Char>> -> RegEx, normalize_range, 1;
    axm %regex.not:                       RegEx -> RegEx, normalize_not,   1;
    axm %regex.quant(optional,star,plus): RegEx -> RegEx, normalize_quant, 1;    $\label{line:regex-quant}$

    lam %regex.lit(val: Char) = %regex.range (val, val);
    let %regex.cls.d          = %regex.range ('0', '9'); // similar: %regex.cls.w, %regex.cls.W,
    let %regex.cls.D          = %regex.not %regex.cls.d; //          %regex.cls.s, %regex.cls.S
\end{lstlisting}
\begin{lstlisting}[label=lst:ptrn,caption=\mim code that constructs RegEx pattern \texttt{\^{}\textbackslash{}w+\textbackslash.[a-z]+\$} (simple top-level domains)\vspace{-2ex},float=tp,abovecaptionskip=-1.5ex,belowcaptionskip=-.5ex]
plugin regex; // parse "regex.mim" and load "libmim.so"
let pattern = %regex.conj (%regex.quant.plus %regex.cls.w,             // '\w+\.[a-z]+'
                           %regex.lit '.', %regex.quant.plus (%regex.range ('a','z')));
\end{lstlisting}
\begin{lstlisting}[language=C++,literate=,label=lst:regex-match,caption={C++ code that matches \texttt{$r$**} and yields \texttt{$r$*}\vspace{-2ex}},float=tp,abovecaptionskip=-1.5ex,belowcaptionskip=-.5ex]
    if (auto star_outer = match(regex::quant::star, r))
        if (auto star_inner = match(regex::quant::star, star_outer->arg())) return star_inner;
\end{lstlisting}
\begin{lstlisting}[language=C++,literate=,label=lst:regex-call,caption={C++ code that constructs a curried call \texttt{{\color{ACMBlue}\%regex.range} (a, b) (mem, str, pos)}\vspace{-2ex}},float=tp,abovecaptionskip=-1.5ex,belowcaptionskip=-1ex]
    world.call<regex::range>(Defs{a, b}, Defs{mem, str, pos})
\end{lstlisting}

\subsection{Plugin Declaration}

The \texttt{\%regex} plugin declares the so-called \emph{axioms} \lstinline|%regexconj|, \lstinline|%regex.disj|, etc.
Line \ref{line:regex-quant} declares the RegEx quantifiers \lstinline|%regex.quant.optional|, \lstinline|%regex.quant.star|, \lstinline|%regex.quant.plus|.
In addition to axioms, plugins can provide supplementary definitions (variables, functions, \ldots) that are entirely defined in \mim.
For example, the digit character class \hl{\%regex.cls.d} is just \lstinline|let|-bound to \hl{\%regex.range ('0', '9')}.
Axioms that inhabit a function type, are usable like ordinary functions.
However, axioms do \emph{not} provide an implementation in \mimir per se.
Their purpose is to denote domain-specific language constructs that the plugin's code transformations refine.
For example, the C++ code in \autoref{lst:regex-match} matches
\mhl{\%regex.quant.star (\%regex.quant.star regex)}
and peels off the superfluous outer quantifier.
Note that \hl{regex::quant::star} stems from the auto-generated header and references \hl{\%regex.quant.star} from \verb|regex.mim|.
Similarly, plugin developers can access or call annexes from C++~(\autoref{lst:regex-call}).

\subsection{Types}
\label{sec:overview_types}

One of \mimir's most prominent features is that it does \emph{not} have a special syntactic category for types.
Instead, types are also expressions.
This has several advantages as we will outline in the following.

First, \mimir does not need special constructs to declare type aliases or type-level functions.
An ordinary \lstinline|let|-binding (line \ref{line:regex-Char} in \autoref{lst:regex-plugin}) defines the alias \hl{Char} for the 8-bit-wide integer type \lst{I8}.
\hl{Str} is also a normal function.
It expects a size \hl{n} of type \lstinline|Nat| and returns~\verb|*|: the type of all types.
Thus, the type of \hl{Str} is \hl{\key{Nat} $\to$ *}.
Here, \hl{Str} yields a pointer to an array of size \hl{n} and element type \hl{Char}.
The pointer type constructor is an axiom from the \texttt{\%mem} plugin: \mhl{axm \%mem.Ptr: * $\to$ *} (\autoref{sec:mem_plugin}).
The \emph{expression} \hl{\arr{e$_n$; T}} introduces an array \emph{type} while both \hl{e$_n$} and \hl{T} are again expressions.
We just use the metavariable \hl{T} to suggest that this is a type expression.
In particular, the size \hl{e$_n$} can be an arbitrarily complex expression and may not necessarily be a compile-time constant as opposed to \hl{n} in C++'s \hl{std::array<T, n>}.
For this reason, \hl{\arr{e$_n$; T}} is called a \emph{dependent type}, as the type depends on a value.
A common misconception is to think of \hl{\arr{e$_n$; T}} as a pair consisting of the size \hl{e$_n$} and the \enquote{actual} array.
This is \emph{not} the case.
It is just an array which \enquote{tracks} its size \hl{e$_n$}.
Dependent types are mostly known from theorem provers such as Coq or Lean.
However, in \mimir we are not concerned about proofs.
We are using dependent types to abstract from the size of integer operations or arrays which in turn allows for type-safe variadic functions and polymorphism over array rank while tracking this dependency in the type system (\autoref{sec:eval}).

The expression \hl{(e$_0$, $\ldots$, e$_{n-1}$)} forms a tuple \emph{value} while \hl{[T$_0$, $\ldots$, T$_{n-1}$]} forms a tuple \emph{type}.
Hence, the function~\hl{Res} returns a tuple \emph{type} which models the result type of a \ac{RegEx} match.
The first element type \hl{\%mem.M} stems again from the \texttt{\%mem} plugin and abstracts from the machine state;
any operation that potentially has a side-effect such as memory accesses consumes a machine state, i.e.~a value of type \hl{\%mem.M}, and produces a new one.
This is similar to the \hl{IO} monad in Haskell.
The second element type \lstinline|Bool| indicates the success or failure of a match.
The last element type \mhl{\key{Idx} n} is an integer within the range \hl{0$_n$}, \ldots, \hl{(n-1)$_n$} and keeps track of the current position within the string to match.
Note that this index type guarantees to access the string in bounds.

The type \hl{RegEx} of a \ac{RegEx} matcher is a \emph{dependent} function type.
\mim allows for convenient specification of curried function types with complex dependencies via \mhl{d $\cdots$ d $\to$ T}:
Each domain~\hl{d} constitutes a curried domain and may as well as the final codomain \hl{T} depend on the \emph{value} of the preceding domains.
Here, the second domain and the codomain \hl{Res n} depend on \hl{n} which is a \emph{value} of the first domain~\lstinline|Nat|.
This dependency makes the first domain deducible when calling a function of type \hl{RegEx}.
This is why first the domain is put within curly braces which marks it as \emph{implicit}:
\mimir will automatically infer this argument when calling a function of type \hl{RegEx}.
The second domain constitutes the \enquote{actual} parameters of the matcher: a machine state, a string, and the current position within this string.

The \hl{RegEx} constructors yield matchers of type \hl{RegEx} while potentially composing other matchers.
For example, \hl{\%regex.not} expects another matcher to negate and \hl{\%regex.range} a range given by two \mbox{\hl{Char}s} as showcased in \autoref{lst:regex-call}:
It creates a range pattern and matches it on a \mbox{\hl{str}ing}.
Note that the size argument~\hl{n} is implicit and inferred in both \mimir's textual representation as well as the C++ code.
The junctions \hl{\%regex.conj} and \hl{\%regex.disj} demonstrate how dependent arrays allow for variadic functions as they expect \hl{i}-many \hl{RegEx} matchers as inputs.

The \texttt{\%regex} plugin does not have to provide any kind of additional validators as would be necessary in \mlir.
All declared annexes have a proper type and \mimir type-checks expressions containing annexes just like any other expression.

\subsection{Normalization}
\label{sec:overview:norm}

Whenever \mimir creates an expression, it is immediately \emph{normalized} (\autoref{sec:norm}).
For example, the tuple extraction \hl{(0, 1, 2)\#2$_3$} is right away resolved to \hl{2}.
In addition, \mimir consistently removes 1-tuples and 1-tuple types.
Hence, it does not matter whether a function is specified as \hl{\key{lam} Str(n: \key{Nat})} or \hl{\key{lam} Str n: \key{Nat}} and whether it is invoked with \hl{Str n} or \hl{Str (n)}.

Normalizations are the backbone of \mimir's optimizer but also influence type checking.
\mimir handles tuples and arrays in a uniform way by normalizing a tuple type \hl{[\key{Nat}, \key{Nat}]} to an array \hl{\arr{2; \key{Nat}}}.
This is why \mimir does not need different introduction and elimination constructs for tuples and arrays:
The type of \hl{(0, 1)} is \hl{\arr{2; \key{Nat}}} and of \hl{(0, \key{ff})} is \hl{[\key{Nat}, \key{Bool}]};
extraction \hl{e\#e$_i$} works regardless of whether \hl{e} is an array or a tuple (\autoref{sec:tuples}).
Thus, \hl{\%regexconj (re1, re2, re3)} types fine:
\mimir infers \hl{3} for \hl{i}, types the argument as \hl{\arr{3; RegEx}}, and removes superfluous 1-tuple types from \hl{\%regex.conj}'s domains.
Additionally, less syntax also implies fewer patterns to match when writing program analyses.

Finally, all expressions are \emph{hash-consed}: % hashed-consed or hash-consed
Whenever an expression is created, \mimir first checks whether a syntactically equal expression already exists.
If this is the case, \mimir will reuse this existing expression.
This has the effect that in the C++ implementation two pointers to \mimir expressions enjoy pointer equality if they are syntactically equal.

Axioms can provide their own \emph{normalizers}: local transformations that are considered generally useful.
The \texttt{\%regex} plugin, for instance, merges quantifiers---\verb|r*?|, \verb|r?*|, \verb|r+?|, \verb|r?+|, \verb|r*+|, \verb|r+*| all normalize to \verb|r*|---and removes idempotence---\verb|r??| results in \verb|r?|, ditto for \verb|*| and \verb|+|.
As the axiom declaration indicates (line \ref{line:regex-quant}), the C++ function \texttt{normalize\_quant}, which is part of \verb|libmim_regex.so|, is the \emph{normalizer} of this axiom and implements this logic.
The actual implementation consists of $\sim\!\!20$ lines of C++ code and involves a few matches and building new calls similar to \autoref{lst:regex-match} and \ref{lst:regex-call}.
The \lstinline|%regex| plugin implements similar normalizations for the other axioms to compute a (pseudo) normal form for \acp{RegEx}.

By default, \mimir fires the specified normalizer when the last curried argument is applied to an axiom and this is in most cases the desired behavior.
However, plugin designers can override this behavior and specify when exactly normalization should happen.
The \texttt{\%regex} plugin is an exception to the default behavior as it wants to normalize, for example, a quantifier as soon as its matcher is applied: \hl{\%regex.quant.star regex}.
If we waited until all curried arguments were passed, we would entirely miss normalization in some instances or be too late for others:
\begin{lstlisting}
    %regex.quant.star (%regex.quant.star regex /*miss*/) (mem, str, pos) /*too late*/
\end{lstlisting}
To this end, the axiom demands normalization when the first argument is passed to the axiom (last part of line \ref{line:regex-quant}).
The same counting mechanism applies when \mbox{\hl{match}ing} curried axiom calls in C++ and is the reason why \autoref{lst:regex-match} works as intended.
Other plugins implement constant folding and various peephole optimizations such as $\hl{x+0}\vtr \hl{x}$ via normalizers.

\subsection{Lowering}
\label{sec:lowering}

Since axioms are opaque entities without implementation, plugins must somehow provide an implementation.
The \texttt{\%mem} plugin contains an \llvm backend that additionally understands the \texttt{\%core} plugin (for integer operations) and \texttt{\%math} plugin (for floating-point operations).
Unless other plugins want to ship their own or extend the existing \llvm backend, they need a phase that substitutes axioms unknown to the \llvm backend to low-level code known to the backend (i.e., only using operations from \texttt{\%mem}, \texttt{\%core}, and \texttt{\%math}).
What is more, many plugins want to apply domain-specific transformations on the code in addition to normalizations before lowering its domain-specific axioms.
To this end, \mimir provides a sophisticated optimizer and a flexible, modular pass manager.
Even compilation phases are exposed as axioms with the help of the \texttt{\%compile} plugin.
This allows users to compose their own compilation pipeline as a \mimir program.
However, the details of the optimizer are beyond the scope of this paper.

As outlined above, the \texttt{\%regex} plugin applies various normalizations to given \acp{RegEx}.
Additionally, the plugin provides a pass written in C++ that constructs \iac{NFA} from a pattern, makes it deterministic, and minimizes it.
Finally, the pass will generate low-level code that implements the minimized \ac{DFA} and replaces the axiom calls with it.
This code is amenable to the \llvm backend.
We discuss the performance of the plugin in \autoref{sec:eval:regex}.

\subsection[Discussion: MimIR vs. Thorin and MLIR]{Discussion: \mimir vs.~\thorin and \mlir}\label{sec:overview:discussion}
\label{sec:mimir_vs_thorin}

\mimir's code base was initially derived from \thorin~\cite{DBLP:conf/cgo/LeissaKH15} but \mimir is now entirely different from its predecessor:
\thorin uses \ac{CPS} for \emph{all} of its functions.
\mimir on the other hand, supports both direct style and \ac{CPS} by giving a continuation the type $T \to \bot$~(see \autoref{sec:fun}).
\mimir adds (higher-order) polymorphism and dependent types with local type inference.
\thorin has a set of hard-coded, second-class built-ins in direct style (much like instructions in \llvm) that \mimir completely removes.
Instead, \mimir provides a plugin system that allows developers to declare not only custom operations but also custom types and type constructors through first-class axioms with a possibly polymorphic, polytypic, and/or dependent type.
In addition, \mimir introduces the \mim language, which allows developers to specify the plugin interface or complete programs directly in \mim.
\mimir's core calculus is now so different from \thorin's, and \mimir's sources hardly contain any \thorin-derived code anymore, that we have renamed the project rather than bumping \thorin's version number.

While \mlir~\cite{DBLP:conf/cgo/LattnerAB0DPRSV21} and \mimir are both extensible compiler frameworks and pursue similar goals,
\mlir only provides a basic infrastructure for hosting languages that has \enquote{little builtin, everything customizable}~\cite[p. 3]{DBLP:conf/cgo/LattnerAB0DPRSV21}. % That's a quote from the \mlir paper. "Little Builtin, Everything Customizable" you mean?
% my problem with this is the dangling "that"... does it refer to the basic infra or the hosted languages?
\mimir, on the other hand, aims to provide a general base language with an expressive type system.
Similarly to \mimir's plugins, \mlir offers extensibility through so-called \emph{dialects}.
Dialects enhance \mlir's parser and type checker because \mlir lacks polymorphism or dependent types.
This raises the question of how the different type-systems of the individual dialects interact and integrate.
There are even dialects such as \circt~\cite{DBLP:journals/corr/abs-2404-18756} that violate basic SSA invariants (no cycles in the data dependence graph).

Additionally, \mlir has only limited support for higher-order functions:
Ops in \mlir do not have a function type and cannot be passed to other functions as first-class citizens.
Custom functions cannot have free variables---like in C.
Nor are regions true functions as regions do not even have a type per se.
They can be syntactically passed to Ops that are specifically designed to expect regions.
Then, the C++ validator checks whether the passed region is used correctly in this specific spot.
But you cannot capture this closure as value and pass it around in \mlir out of the box.
The \verb|lp| dialect~\cite{DBLP:conf/cgo/BhatG22} uses this feature to implement full closure support but \lstinline|lp| works on a type-erased representation.
For example, the types of all higher-order arguments have been erased to \lstinline|!lp.t|---a boxed heap value.

% vim:spell:spelllang=en
\section{Semantics}
\label{sec:sema}

In the following, we first present the formal definition of \mimir including its syntax, semantics, and normalization rules (\autoref{fig:rules}).
\mim supports syntactic sugar, which \mim translates into the core syntax (\autoref{fig:sugar}).
In order to keep this presentation as concise as possible, we leave out a few details and full recursion which we will discuss afterward.
Then, we introduce \mimir's partial evaluator that tightly interacts with type checking and normalization.
Finally, we present \mimir's type safety.

\subsection{Syntax, Typing \& Normalization}
\label{sec:sema:intro}

\paragraph{Preliminaries}

Since \mimir is based on \ac{PTS}, it uses the same syntax for terms, types, and kinds.
However, we usually use the metavariable \hl{e} to evoke a term expression and \hl{T} or \hl{U} to evoke a type expression.
\mimir uses a stratified, countably infinite hierarchy of sorts which are organized in a predicative way (\irefrule{Sort}).
This avoids well-known paradoxes (like Girard's paradox) associated with self-referential definitions.
As syntactic sugar we write \hlstar{} for \hl{\key{Sort}\ 0}: the type of all types.
Type constructors such as \hl{* $\to$ *} are of type \hl{\key{Sort}\ 1} etc.

\mimir knows several \emph{binders}.
These are expressions that introduce a variable of the form~\hl{x:e}.
We follow \emph{Barendregt’s convention}:
No variable is both free and bound;
every bound variable is bound \emph{exactly once}.
While \mim supports lexical scoping, \mimir's graph representation (\autoref{sec:graph}) actually ensures Barendregt's convention.

We call a binder \emph{parametric}, if the introduced variable occurs free in any subsequent expression of the binder.
Otherwise, we call the binder \emph{non-parametric}.
There is syntactic sugar for non-parametric binders available which omits the variable altogether (\autoref{fig:sugar}).

\def \MathparLineskip {\lineskip=1.5ex}
\begin{figure}[p]
    \centering
    \begin{scriptsize}
        \begin{tabular}{c|c}\hspace{-7ex}
            \begin{minipage}{.65\textwidth}
                \vspace{-1ex}
                \begin{equation*}
                    \arraycolsep=2pt
                    \begin{array}{rll}
                        \hlgamma                        &\!\Coloneqq \cdot \mid \hlgamma,\hl{x}:\hl{T} & \textit{Typing Environment} \\
                        \hul{e},\hul{T},\hul{U}         &\!\Coloneqq \hl{\key{Sort}}\ \hl{s}
                                                                \mid \hl{$\bot$}
                                                                \mid \key{Nat}
                                                                \mid \hl{\key{Idx}}
                                                                \mid \hl{n}
                                                                \mid \hl{i$_\texttt{n}$} & \textit{Sort / Bottom / Nat / Idx / Literal} \\
                                                              & \Mid \hl{x}
                                                                \mid \hl{\key{let} x = \ul{e}\lstinline|;|\ \ul{e}}
                                                                \mid \hl{\key{axm} x:\ul{T}; \ul{e}} & \textit{Var / Let / Axiom \ \quad\qquad $\hl{i},\hl{n},\hl{s} \in \mathbb{N}$} \\
                                                              & \Mid \hl{[x:\ul{T}] $\to$ \ul{U}}
                                                                \mid \hl{$\lambda$x:\ul{T}@\hl{e}:\ul{U} = \ul{e}}
                                                                \mid \hl{\ul{e}\ \ul{e}}    & \textit{Pi / Lam / App \ \,\qquad\qquad $\hl{i} < \hl{n}$} \\
                                                              & \Mid \hl{[}\overline{\hl{x:\ul{T}}}\hl{]}
                                                                \mid \hl{(}\overline{\hul{e}}\hl{)}
                                                                \mid \hl{\arr{x:\ul{e}; \ul{T}}}
                                                                \mid \hl{\pack{x:\ul{e}; \ul{e}}}
                                                                \mid \hl{\ul{e}\#\ul{e}}   & \textit{Sigma / Tuple / Array / Pack / Extract} \\
                    \end{array}
                \end{equation*}
            \end{minipage}
            &
            \begin{minipage}{.29\textwidth}
                \begin{scriptsize}
                    \begin{mathpar}
                        \framebox{$\step{e}{e}$}
                        \and
                        \irule{Cong}{
                            \step{e}{e'}
                        }{
                            \mathcal{E}[{\hl{e}}] \to \mathcal{E}[{\hl{e'}}]
                        }
                        \and
                        \irule{Beta}{
                        }{
                            \hl{($\lambda$ x:T@\hl{e}:U = e$_b$)\ e$_a$} \to \hl{e$_b$}\subst{e$_a$}{x}
                        }
                    \end{mathpar}
                \end{scriptsize}
            \end{minipage}
        \end{tabular} \\[.5ex]
        \noindent\hspace{-.5ex}$%
            \arraycolsep=2pt
            \begin{array}{rll}
                \mathcal{E}[\hl{\hole}]    &\!\Coloneqq \hl{\key{axm} x:T; \hole}
                                                    \mid \hl{\key{axm} x:\hole; e}
                                                    \mid \hl{[\mydots, \hole, \mydots]}
                                                    \mid \hl{(\mydots, \hole, \mydots)}
                                                    \mid \hl{\arr{x:\hole; e}}
                                                    \mid \hl{\arr{x:e; \hole}}
                                                    \mid \hl{\pack{x:\hole; e}}
                                                    \mid \hl{\pack{x:e; \hole}} & \textit{Evaluation\phantom{xxxxxxxxxxxx}}\\
                                                &   \Mid \hl{\hole\ e}
                                                    \mid \hl{e\ \hole}
                                                    \mid \hl{\hole\,\#e}
                                                    \mid \hl{e\#\,\hole}
                                                    \mid \hl{[x:\hole] $\to$     U}
                                                    \mid \hl{[x:T]     $\to$ \hole}
                                                    \mid \hl{$\lambda$x:\hole\,@\hl{e}:U = \hl{e}}
                                                    \mid \hl{$\lambda$x:T@\hl{e}:\hole\, = \hl{e}}
                                                    \mid \hl{$\lambda$x:T@\hl{e}:U = \hole} & \textit{\quad{}Context}
            \end{array}$
        \hrule
        \vspace{-1ex}
        \begin{mathpar}
            \framebox{$\hlgamma \vdash \hl{e} : \hl{T}$}
            \and
            \irule{Sort}{
                \hl{n'} = \hl{n} + 1
            }{
                \type{\key{Sort}\ n}{\key{Sort}\ n'}
            }
            \and
            \irule{Bot}{
            }{
                \type{$\bot$}{\hlstar}
            }
            \and
            \irule{Nat}{
            }{
                \type{\key{Nat}}{\hlstar}
            }
            \and
            \irule{Idx}{
            }{
                \type{\key{Idx}}{\key{Nat} $\to$ \hlstar}
            }
            \and
            \irule{Lit-N}{
            }{
                \type{n}{\key{Nat}}
            }
            \and
            \irule{Lit-I}{
                \hl{i} < \hl{n}
            }{
                \type{i$_n$}{\key{Idx} n}
            }
            \and
            \irule{Var}{
                \hl{x}:\hl{T} \in \hlgamma
            }{
                \type{x}{T}
            }
            \and
            \irule{Ax}{
                \hlgamma \vdash \hl{T} : \hl{\key{Sort}\ s} \\
                \hlgamma,\hl{x} : \hl{T} \vdash \hl{e} : \hl{U}
            }{
                \type{\key{axm} x : T; e}{U}
            }
            \and
            \irule{Pi}{
                \type{T}{\hl{\key{Sort}\ s$_t$}} \\
                \hlgamma,\hl{x}:\hl{T} \vdash \hl{U} : \hl{\key{Sort}\ s$_u$}
            }{
                \hlgamma \vdash \hl{[x:T] $\to$ U} : \hl{\key{Sort}}\ \max\{\hl{s$_t$},\hl{s$_u$}\}
            }
            \and
            \irule{Lam}{
                \hlgamma,\hl{x}:\hl{T} \vdash \hl{e$_f$} : \key{Bool} \\
                \hlgamma,\hl{x}:\hl{T} \vdash \hl{U} \rto \hl{e} \\
                \type{[x:T] $\to$ U}{\key{Sort} s}
            }{
                \type{$\lambda$ x:T@e$_f$: U = e}{[x:T] $\to$ U}
            }
            \and
            \irule{App}{
                \type{e}{[x:T] $\to$ U} \\
                \atype{T}{e$_T$}
            }{
                \type{e e$_T$}{U}[\hl{e$_T$}/\hl{x}]
            }
            \and
            \irule{Sig}{
                \type{T$_0$}{\hl{\key{Sort}\ s$_0$}} \quad
                \cdots \quad
                \hlgamma,\hl{x$_0$}:\hl{T$_0$},\ldots,\hl{x$_{n-2}$}:\hl{T$_{n-2}$} \vdash \hl{T$_{n-1}$} : \hl{\key{Sort}\ s$_{n-1}$}
            }{
                \hlgamma \vdash \hl{[x$_0$:T$_0$,\mydots,x$_{n-1}$:T$_{n-1}$]} : \hl{\key{Sort}}\ \max\{\hl{s$_0$},\mydots,\hl{s$_{n-1}$}\}
            }
            \and
            \irule{Tup}{
                \type{e$_0$}{T$_0$} \quad
                \cdots \quad
                \type{e$_{n-1}$}{T$_{n-1}$} \\\\
                \hl{[T$_0$,\mydots,T$_{n-1}$]} \vtr \hl{T} \\
                \type{T}{\key{Sort} s}
            }{
                \type{(e$_0$,\mydots,e$_{n-1}$)}{T}
            }
            \and
            \irule{Arr}{
                \type{e$_n$}{\key{Nat}} \\\\
                \hlgamma,\hl{x}:\hl{\key{Idx} e$_n$} \vdash \hl{T} : \hl{\key{Sort}\ s}
            }{
                \type{\arr{x: e$_n$; T}}{\key{Sort}\ s}
            }
            \and
            \irule{Pack}{
                \type{e$_n$}{\key{Nat}} \\
                \hlgamma,\hl{x}:\hl{\key{Idx} e$_n$} \vdash \hl{e} : \hl{T} \\\\
                \hl{\arr{x: e$_n$; T}} \vtr \hl{U} \\
                \type{U}{\key{Sort} s}
            }{
                \type{\pack{x: e$_n$; e}}{U}
            }
            \and
            \irulex{Ex-SL}{Ex-S$_L$}{
                \type{e}{[x$_0$:T$_0$,\mydots,x$_{n-1}$:T$_{n-1}$]} \\
                \hl{i} < \hl{n}
            }{
                \hlgamma \vdash \hl{e\#i$_n$} : \hl{T$_i$}\subst{e\#0$_n$}{x$_0$}\cdots\subst{e\#(i-1)$_n$}{x$_{i-1}$}
            }
            \and
            \irulex{Ex-SI}{Ex-S$_i$}{
                \type{e$_i$}{\key{Idx} n} \\
                \type{e}{[x$_0$:T$_0$,\mydots,x$_{n-1}$:T$_{n-1}$]} \\
                \forall_{1 \le i < n}.
                \type{T$_i$}{\key{Sort}\ s} \\\\
                \hl{T$_j$'} = \hl{T$_j$}\subst{e\#0$_n$}{x$_0$}\cdots\subst{e\#(j-1)$_n$}{x$_{j-1}$} \\
                \hl{(T$_0$',\mydots,T$_{n-1}$')} \vtr \hl{T} \\
                \hl{T\#e$_i$} \vtr \hl{U} \\
            }{
                \hlgamma \vdash \hl{e\#e$_i$} : \hl{U}
            }
            \and
            \irule{Ex-A}{
                \type{e}{\arr{x:e$_n$;T}} \\\\
                \type{e$_i$}{\key{Idx} e$_n$}
            }{
                \hlgamma \vdash \hl{e\#e$_i$} : \hl{T}\subst{e$_i$}{x}
            }
        \end{mathpar}
        \vspace{-1ex}
        \hrule
        \vspace{-1ex}
        \begin{mathpar}
            \framebox{$\atype{T}{e}$}
            \and
            \irule{A-T}{
                \type{e}{T}
            }{
                \atype{T}{e}
            }
            \and
            \irule{A-Tup}{
                \atype{T$_0$}{e\#0$_n$} \\
                \forall_{1 \le i < n}.
                \hlgamma \vdash \hl{T$_i$}\subst{e\#0$_n$}{x$_0$}\cdots\subst{e\#({i-1})$_n$}{x$_{i-1}$} \rto \hl{e\#i$_n$} \\
                %\type{[x$_0$:t_0,\mydots,x$_{n-1}:t_{n-1}]}{s}
            }{
                \atype{[x$_0$:T$_0$,\mydots,x$_{n-1}$:T$_{n-1}$]}{e}
            }
        \end{mathpar}
        \vspace{-1ex}
        \hrule
        \vspace{-1ex}
        \begin{mathpar}
            \framebox{$\hul{e} \vtr \hl{e}$}
            \and
            \irule{N-Let}{
            }{
                \hl{\key{let} x = e; e'} \vtr \hl{e'}[\hl{e}/\hl{x}]
            }
            \and
            \irulex{N-Ex1}{N-Ex$_1$}{
            }{
                \hl{e\#0$_1$} \vtr \hl{e}
            }
            \and
            \irulex{N-Tup1}{N-Tup$_1$}{
            }{
                \hl{(e)} \vtr \hl{e}
            }\hspace{-2ex}
            \and
            \irulex{N-Sig1}{N-Sig$_1$}{
            }{
                \hl{[T]} \vtr \hl{T}
            }
            \and
            \irulex{N-Pack1}{N-Pack$_1$}{
            }{
                \hl{\pack{1; e}} \vtr \hl{e}
            }
            \and
            \irulex{N-Arr1}{N-Arr$_1$}{
            }{
                \hl{\arr{1;T}} \vtr \hl{T}
            }
            \and
            \irulex{N-TupB}{N-Tup$_\beta$}{
            }{
                \hl{(e$_0$,\mydots,e$_{n-1}$)\#i$_n$} \vtr \hl{e$_i$}
            }
            \and
            \irulex{N-PackB}{N-Pack$_\beta$}{
            }{
                \hl{\pack{n; e}\#e$_i$} \vtr \hl{e}
            }
            \and
            \irulex{N-TupE}{N-Tup$_\eta$}{
            }{
                \hl{(e\#0$_n$,\mydots,e\#(n-1)$_n$)} \vtr \hl{e}
            }
            \and
            \irule{N-PackTup}{
                \hl{n} > 1
            }{
                \hl{(}\underbrace{\hl{e,\mydots,e}}_{\text{$\hl{n}$ times}}\hl{)} \vtr \hl{\pack{n;e}}
            }
            \and
            \irule{N-ArrSig}{
                \hl{n} > 1
            }{
                \hl{[}\underbrace{\hl{T,\mydots,T}}_{\text{$\hl{n}$ times}}\hl{]} \vtr \hl{ \arr{n;T}}
            }
            \and
            \irulex{N-B}{N-$\beta$}{
                \hl{e$_f$}\subst{e$_a$}{x} \equiv \key{tt}
            }{
                \hl{($\lambda$ x:T@e$_f$:U = e$_b$)\ e$_a$} \vtr \hl{e$_b$}\subst{e$_a$}{x}
            } \vspace{-1ex}\\
            \irule{N-TupPack}{
                \hl{n} \in \mathbb{N} \\
                \hl{x} \in \FV{e}
            }{
                \hl{\pack{x:n;e}}
                \vtr
                \hl{(e}\subst{0$_n$}{x}\hl{,\mydots,e}\subst{(n-1)$_n$}{x}\hl{)}
            }
            \and
            \irule{N-SigArr}{
                \hl{n} \in \mathbb{N} \\
                \hl{x} \in \FV{e}
            }{
                \hl{\arr{x:n;T}}
                \vtr
                \hl{[T}\subst{0$_n$}{x}\hl{,\mydots,T}\subst{(n-1)$_n$}{x}\hl{]}
            }
            \and
            \irule{N-Id}{
                \text{otherwise}
            }{
                \hul{e} \vtr \hl{e}
            }
        \end{mathpar}
        \hrule
        \vspace{-3.5ex}
    \end{scriptsize}
    \caption{Syntax, $\beta$-Reduction, Typing, \& Normalization. $\hul{e}$ \emph{might not} be normalized; $\hl{e}$ \emph{is} normalized.}
    \label{fig:rules}
    \vspace{1ex}
    \hrule
    \vspace{.5ex}
    \hrule
    \vspace{.5ex}
    \begin{scriptsize}
        \begin{subcaptionblock}{\textwidth}
            {\scriptsize{}Sort/Integer\hfill{}Bool\hfill{}non-parametric binders\hfill{}function/continuation type}\\[-4ex]
            \begin{align*}
                \hl{*}            &\coloneqq \hl{\key{Sort}    0}   & \hl{\key{Bool}}    &\coloneqq \hl{\key{Idx} 2} & \hl{[}\mydots\hl{,T,}\mydots\hl{]}&\coloneqq \hl{[}\mydots\hl{,\textunderscore:T,}\mydots\hl{]}  & \hl{T $\to$ U}                     &\coloneqq \hl{[\textunderscore:T] $\to$ U} \\[-1ex]
                \hl{\key{I8}}    &\coloneqq \hl{\key{Idx} 0x100}   &  \hl{\key{ff}}      &\coloneqq \hl{0$_2$}        & \hl{\arr{e$_n$;T}}              &\coloneqq \hl{\arr{\textunderscore:e$_n$;T}}                & \hl{\key{Cn} T \phantom{$\to$ U}} &\coloneqq \hl{T $\to$ $\bot$}                  \\[-1ex]
                \hl{\key{I16}}   &\coloneqq \hl{\key{Idx} 0x10000}\ \text{etc.} &\hl{\key{tt}}      &\coloneqq \hl{1$_2$}        & \hl{\pack{e$_n$;e}}             &\coloneqq \hl{\pack{\textunderscore:e$_n$;e}}               & \hl{\key{Fn} T $\to$ U}           &\coloneqq \hl{\key{Cn} [T, \key{Cn} U]}
            \end{align*}
            \\[-4ex]
            \noindent{\scriptsize{}where}\\[-6.5ex]
            \begin{align*}
                \hl{e \key{where} \key{let} x1 = $\mydots$; $\cdots$ \key{let} xn = $\mydots$; \key{end}} \coloneqq \hl{\key{let} x1 = $\mydots$; $\cdots$ \key{let} xn = $\mydots$; e}
            \end{align*}
            \\[-4ex]
            \noindent{\scriptsize{}anonymous function/continuation\hfill{}named function/continuation}\\[-4ex]
            \begin{align*}
                \hl{$\lambda$ x:T :U = e}             &\coloneqq \hl{$\lambda$ x:T@\key{tt}:U = e}          & \hl{\key{lam} f x:T :U = e}           &\coloneqq \hl{\key{let} f = \ \  $\lambda$ x:T :U = e} \\[-1ex]
                \hl{\key{cn} x:T \phantom{:U} = e}   &\coloneqq \hl{$\lambda$ x:T@\key{ff}:$\bot$\! = e}     & \hl{\key{con} f x:T \phantom{:U} = e} &\coloneqq \hl{\key{let} f = \key{cn} x:T \phantom{:U} = e}   \\[-1ex]
                \hl{\key{fn} x:T :U = e}             &\coloneqq \hl{\key{cn} (x:T, return:\key{Cn} U) = e} & \hl{\key{fun} f x:T :U = e}           &\coloneqq \hl{\key{let} f = \key{fn} x:T :U = e}
            \end{align*}
            \vspace{-4ex}
            \caption{Simple syntactic sugar}
            \label{fig:simple_sugar}
        \end{subcaptionblock}
        \\[2ex]

        \begin{subcaptionblock}{.62\textwidth}
            \begin{lstlisting}[basicstyle=\scriptsize\ttfamily]
                LAM  (T: *) ((x y: T), return: T -> BOT): BOT = return x
                cn (T: *) ((x y: T), return: Cn T)     = return x
                fn (T: *)  (x y: T): T                 = return x
                LAM T: * @tt: [_: [[T, T], [_: T] -> BOT]] -> BOT =
                    LAM xyr: [[T, T], [_: T] -> BOT]@ff: BOT = xyr#1_2 xyr#0_2#0_2
            \end{lstlisting}
            \vspace{-1.ex}
            \caption{Curried functions/continuations}
            \label{lst:fun_expr}
        \end{subcaptionblock}%
        \begin{subcaptionblock}{.38\textwidth}
            \begin{lstlisting}[basicstyle=\scriptsize\ttfamily,gobble=16]
                   [T: *] [[T, T], T -> BOT] -> BOT
                Cn [T: *] [[T, T], Cn T]
                Fn [T: *]  [T, T] -> T

                [T:*] -> [_:[[T, T], [_:T] -> BOT]] -> BOT
            \end{lstlisting}
            \vspace{-1.ex}
            \caption{Curried function/continuation types}
            \label{lst:fun_type}
        \end{subcaptionblock}
    \end{scriptsize}
    \vspace{-3ex}
    \caption{Syntactic sugar (excerpt).
        All functions in \ref{lst:fun_expr} are equivalent.
        The type can be expressed by any expression in \ref{lst:fun_type}---they are equivalent, too.
        The last respective item depicts the completely desugared version.
        Similar sugar is available for \lstinline|lam|, \lstinline|con|, \lstinline|fun|.
        In addition, \lstinline|lam|/\lstinline|con|/\lstinline|fun| allow for recursion (\autoref{sec:rec}).}
    \label{fig:sugar}
\end{figure}

%{\scriptsize\hfill{}curried function/continuation type\\[-1ex]\hfill\lstinline|Fn| similar}\\[-8ex]
%\begin{flalign*}
    %\hl{$\Pi$ [a: A, b: B a][x: X a, y: Y b x] $\to$ F b x} &\coloneqq \\
        %\intertext{\hfill\hl{$\Pi$ ab:[a: A, B a] $\to$ $\Pi$ xy:[x: X ab\#0$_2$, y: Y ab\#1$_2$ x] $\to$ F ab\#1$_2$ xy\#0$_2$}}
    %\hl{\key{Cn} [a: A, b: B a][x: X a, y: Y b x]} &\coloneqq \hl{$\Pi$ ab:[a: A, B a] $\to$ \key{Cn} xy:[x: X ab\#0$_2$, y: Y ab\#1$_2$ x]}
%\end{flalign*}
%{\scriptsize\hfill{}curried function/continuation\\[-1ex]\hfill\lstinline|fn|, \lstinline|lam|, \lstinline|con|, \lstinline|fun| similar}\\[-8ex]
%\begin{multline*}
    %\hl{$\lambda$ (a: A, b: B a)(x: X a, y: Y b x): F b x = f b x} \coloneqq \\
        %\hl{$\lambda$ ab:[a: A, B a]: $\Pi$[x: X ab\#0$_2$, Y ab\#1$_2$ x] $\to$ F ab\#1$_2$ x = } \\
            %\shoveright{\hl{$\lambda$ xy:[x: X ab\#0$_2$, Y ab\#1$_2$ x]: F ab\#1$_2$ xy\#0$_2$ = f ab\#1$_2$ xy\#0$_2$}} \\
    %\shoveleft{\hl{\key{cn} (a: A, b: B a)(x: X a, y: Y b x) = k b x} \coloneqq} \\
        %\hl{$\lambda$ ab:[a: A, B a]: \key{Cn} [x: X ab\#0$_2$, Y ab\#1$_2$ x] =} \\
            %\hl{\key{cn} xy:[x: X ab\#0$_2$, Y ab\#1$_2$ x] = k ab\#1$_2$ xy\#0$_2$}
%\end{multline*}

\subsubsection{Normalization}
\label{sec:norm}

Whenever \mimir builds an expression, it will immediately \emph{normalize} it according to $\hul{e} \vtr \hl{e}$.
Normalization rules play not only an important role in \mimir's optimizer but also in its type checker.
First, types themselves are normalized.
Second, a normalized expression may appear as an argument to another type constructor.
In particular in the case of dependent types, checking for type equality requires checking for program equivalence (see also \autoref{sec:sema_pe}).
\begin{example}
    Consider the function:
    \begin{lstlisting}
        LAM (a: <<%core.nat.add (0, n); T>>): U = body
    \end{lstlisting}
    The \texttt{\%core} plugin (\autoref{sec:case:ir}) normalizes the addition and, hence, simplifies the function to:
    \begin{lstlisting}
        LAM (a: <<n; T>>): U = body
    \end{lstlisting}
    This allows a caller to pass a value of type \hl{<<n; T>>} to this function.
    This would be ill-typed without normalization.
\end{example}

Most dependently typed languages suffer from artifacts such as \hl{f (n + 0)} types in some context whereas \hl{f (0 + n)} does not.
\mimir's extensible normalization framework is able to mitigate such issues.
\begin{example}
    The \texttt{\%core} plugin normalizes both
    \begin{equation*}
        \hl{f (\%core.nat.add (0, n))} \qtext{and} \hl{f (\%core.nat.add (n, 0))} \qtext{to} \hl{f n}\ .
    \end{equation*}
\end{example}

We \emph{underline} an expression $\hul{e}$ to denote that it \emph{might not} be \emph{normalized}.
A \emph{non-underlined} expression $\hl{e}$ denotes that it \emph{is} already \emph{normalized}.
Non-normalized expressions only exist in \mim which is compiled into \mimir's internal graph form early in the compilation pipeline.
\mim translates an expression to \mimir by \emph{first} recursively translating all subexpressions into the desugared, normalized \mimir graph representation, and \emph{then} desugaring the current expression and assembling it via the $\vtr$-relation.
Therefore, only normalized expressions exist within \mimir itself.
For this reason, the normalization rules do not include premises to normalize subexpressions as \mimir will only need to assemble expressions from subexpressions that \emph{are} already normalized.
\mim only performs rudimentary semantic checks such as name analysis.
Type checking happens on \mimir's normalized graph representation.
\begin{example}
    \mimir will never need to normalize \hl{((e))}.
    As soon as \mim translates \hl{(e)} to its internal graph, \irefrulex{N-Tup1}{N-Tup$_1$} fires and yields \hl{e}.
    Then, the outer parentheses form \hl{(e)} again.
    This will fire \irefx{N-Tup1}{N-Tup$_1$} once more and yield \hl{e}.
\end{example}
\begin{example}
    Consider \hl{\key{let} x = 3; x}.
    \irefrule{N-Let} will immediately normalize this expression to~\hl{3}.
    In fact, \lstinline|let|-expressions \emph{only} exist in \mim as \iref{N-Let} will eliminate them outright.
    For this reason, there is neither a typing nor an evaluation rule for a \lstinline|let|-expression.
    Note that all expressions are hash-consed (\autoref{sec:graph}) and, thus, appear exactly once in the program graph.
    In fact, the implementation does not really perform a substitution here.
    The front-end just memorizes that \hl{x} refers to a certain subgraph and all uses of \hl{x} will be wired to that subgraph.
\end{example}

Plugins have the opportunity to extend normalizations (see \autoref{sec:overview:norm}) by adding rules of the form:
$\hl{\%myaxmiom\ e$_1$\ $\cdots$\ e$_n$} \triangleright \hl{e$_\mathit{something}$}$.
Typical examples include constant folding or various identities like $\hl{x+0}\vtr \hl{x}$.

Normalizations must be deterministic and cycle-free.
Cyclic rules such as
$\hl{x + x} \vtr \hl{2 $\cdot$ x}$
and
$\hl{2 $\cdot$ x}  \vtr  \hl{x + x}$
can cause \mimir to diverge as it may endlessly oscillate between these two rules.
Determinism is achieved automatically since rules are implemented in C++ which is executed deterministically.
However, in the future, we would like to permit the plugin designer to directly specify rewrite rules in \mim.
This would allow plugin designers to specify nondeterministic rules but \mimir could also issue warnings or errors in the case of potentially nondeterministic or cyclic rules.

\paragraph{Substitution}

Substitution $\hl{e}\subst{e$_b$}{e$_a$} = \hl{e$_s$}$ where $\hl{e$_a$}$ is recursively replaced with \hl{e$_b$} within \hl{e} is defined in the usual manner with the following extension:
\emph{All expressions created along the way are also normalized}.
Thus, the resulting expression $\hl{e$_s$}$ is again normalized.
Note that \hl{e}, \hl{e$_b$}, and \hl{e$_a$} are also normalized.

\begin{example}\label{ex:subst}
    The substitution $\hl{e\#x}\subst{$0_1$}{x}$ directly yields $\hl{e}$ as \iref{N-Ex1} removes the superfluous extraction when the substitution assembles all substituted subexpressions.
\end{example}

\subsubsection{Nat \& Idx}

\mimir has a builtin type \lstinline|Nat| inhabited by \hl{0}, \hl{1}, \hl{2}, \ldots.
Given \hl{e} of type \lstinline|Nat|, the type \hl{\key{Idx} e} represents integers within the range \hl{0$_e$}, \ldots, \hl{(e-1)$_e$} (\iref{Nat}/\iref{Idx}/\iref{Lit-N}/\iref{Lit-I}).
For example, type \hl{\key{Idx} 3} has three inhabitants: \hl{0$_3$}, \hl{1$_3$}, \hl{2$_3$}.
For convenience, \lstinline|Bool| is an alias for \mbox{\hl{\key{Idx} 2}}, as this type has exactly two inhabitants, for whom appropriate aliases are available, too: $\lstinline|ff| \coloneqq \hl{0$_2$}$ (false) and $\lstinline|tt| \coloneqq \hl{1$_2$}$ (true).
In addition, there is \lstinline|I8|, \lstinline|I16|, \ldots available for \hl{\key{Idx}\ 0x100}, \hl{\key{Idx}\ 0x10000}, \ldots.
We use these types for bootstrapping axioms, but they also play an important role within \mimir itself:
The \emph{arity}, i.e.~the number of elements of a tuple/array, is a value of type \lstinline|Nat|, whereas a value of type \hl{\key{Idx} e} addresses a specific element of a tuple/array with arity~\hl{e}.

\subsubsection[Tuples, Packs, Arrays \& Sigma-Types]{Tuples, Packs, Arrays \& $\Sigma$-Types}
\label{sec:tuples}

Most languages distinguish between array and tuple terms as well as their types.
Albeit \mimir does have different syntax, this distinction does not matter much because \mimir normalizes between both representations.

%\paragraph[Sigma-Types]{$\Sigma$-Types}

\mimir generalizes dependent pair types to n-ary dependent tuple types ($\Sigma$-types) where the \emph{type} of an element may depend on the \emph{value} of \emph{any preceding} element~(\iref{Sig}).
A dependent pair---denoted by \hl{$\Sigma$ x:T.U} in literature---is written as \hl{[x: T, U]} in \mimir.
However, \mimir allows for more complex dependent $\Sigma$-types:%
\begin{lstlisting}
    let Num = [T: *, add: [T, T] -> T, mul: [T, T] -> T, _0:  T, _1: T];
\end{lstlisting}

The expression \hl{\arr{x:e$_n$; T}} forms an \emph{array type} with \hl{e$_n$}-many elements and element type~\hl{T}.
The \emph{arity} \hl{e$_n$} must be of type \lstinline|Nat| and may introduce a variable~\hl{x} whose type is \hl{\key{Idx} e$_n$} and may be used inside the \emph{body}~\hl{T} (\iref{Arr}).
\irefrule{N-ArrSig} expands \emph{parametric} arrays of \emph{constant} arity to tuple types whereas \irefrule{N-SigArr} compresses \emph{homogeneous} tuple types to arrays.
This compression is not strictly necessary but makes the implementation more efficient.
In particular, huge arrays like \hl{\arr{1000000; T}} are not expanded at all.

The term \hl{(e$_0$, $\mydots$, e$_{n-1}$)} introduces a \emph{tuple} while \hl{\pack{x:e$_n$; e}} introduces a \emph{pack}.
Think of a pack as a compressed tuple.
Both terms either inhabit a $\Sigma$-type or an array.
Tuples and packs work analogously to tuple types and arrays but on term level (\iref{Tup}/\iref{Pack}/\iref{N-TupPack}/\iref{N-PackTup}).
\begin{example}
    Due to normalization \mimir considers both the non-normalized as well as the normalized expressions as equal:
    \begin{align*}
        \hl{[\key{Nat}, \key{Nat}]} &\vtr \hl{\arr{2;\key{Nat}}} &
        \hl{\arr{i:2; F\ i}} &\vtr \hl{[F\ 0$_2$, F\ 1$_2$]} \\
        \hl{(0, 0)} &\vtr \hl{\pack{2;0}} &
        \hl{\pack{i:2; f\ i}} &\vtr \hl{(f\ 0$_2$, f\ 1$_2$)}
    \end{align*}
\end{example}

The term \hl{e\#e$_i$} \emph{extracts} from \hl{e} the element with index \hl{e$_i$}.
This index must type as \hl{\key{Idx} e$_n$}.
If \hl{e} types as array with \hl{e$_n$}-many elements, the type of the extraction is the body of the array while substituting the array's variable with the given index~\hl{e$_i$}~(\iref{Ex-A}).
If \hl{e} types as $\Sigma$-type with \hl{e$_n$}-many elements, there are two subcases.
\irefx{Ex-SL}{Ex-S$_L$} picks the \hl{i}$^\mathit{th}$ element type (while substituting all preceding variables~\hl{x$_j$} with \hl{e\#j$_n$}), if the index is a literal~\hl{i$_n$}.
\begin{example}
    Suppose \hl{nmx} has type \hl{[n: Nat, m: Nat, x: F n m]}.
    Then, \hl{nmx\#2$_3$} has type \mhl{F nmx\#0$_3$ nmx\#1$_3$}.
\end{example}
\begin{wraptable}{r}{.35\textwidth}
    \centering
    \vspace{-2ex}
    \begin{footnotesize}
        \begin{tabular}{ll}
            \toprule
            Expression              & Type \\
            \midrule
            \lstinline|(0, 1, 2)|         & \lstinline|<<3; Nat>>| \\
            \lstinline|(0, 1, 2)#i|       & \lstinline|Nat| \\
            \lstinline|(0, tt)|          & \lstinline|[Nat, Bool]| \\
            \lstinline|(0, tt)#0_2|      & \lstinline|Nat| \\
            % \lstinline|(0, tt)#1_2|      & \lstinline|Bool| \\
            \lstinline|(0, tt)#i|        & \lstinline|(Nat, Bool)#i| \\
            \lstinline|(Nat, Bool)#i|   & \lstinline|<2; *>#i| $\vtr$ \lstinline|*| \\
            \lstinline|(0, Bool)#i|      & $\lightning$ \\
            \bottomrule
        \end{tabular}
        \caption{Typing examples}\label{tab:types}
        \vspace{-5ex}
    \end{footnotesize}
\end{wraptable}
If the index is not a literal, \irefx{Ex-SI}{Ex-S$_i$} types the extraction as another extraction from a tuple \enquote{one level up}.
Each element of this tuple contains the corresponding element type of the $\Sigma$-type (while substituting all preceding variables~\hl{x$_j$} with \hl{e\#j$_n$}).
This is only allowed, if all involved element types agree on the same sort (see \autoref{ex:arr}).

\irefrulex{N-B}{N-$\beta$} eliminates a tuple extraction with a known index while \irefx{N-PackB}{N-Pack$_\beta$} eliminates an extraction---no matter the index---from a non-parametric pack.
\irefrulex{N-TupE}{N-Tup$_\eta$} resolves a tuple comprised of a sequence of extractions from the same entity with increasing indices.
Finally, \mimir consistently removes 1-tuples/packs (\irefx{N-Tup1}{N-Tup$_1$}, \irefx{N-Pack1}{N-Pack$_1$}), 1-tuple types/arrays (\irefx{N-Sig1}{N-Sig$_1$}, \irefx{N-Arr}{N-Arr$_1$}), and extractions with \hl{0$_1$} (\irefx{N-Ex1}{N-Ex$_1$}).
\begin{example}\label{ex:arr}
    Note that most languages need different syntax to introduce a tuple or an array term such as \hl{(0, 1, 2)} vs.~\hl{[0, 1, 2]} in Rust.
    \mimir does not need this distinction.
    As \autoref{tab:types} showcases, tuples with homogeneous element types are typed as array.
    This makes them amendable for extractions with an unknown index.
    However, tuples with inhomogeneous element types are typed as $\Sigma$-type.
    Extraction with a known index yields the corresponding element type as expected.
    However, extraction with an unknown index yields a type computation as another extraction (third last row).
    This again yields a type computation as another extraction (second last row).
    If \irefrulex{Ex-SI}{Ex-S$_i$} would not prohibit extraction with an unknown index for a tuple whose elements do not agree on their sort, the type of the expression in the last row would be \hl{(\key{Nat}, \key{Sort} 0)\#i} of type \hl{(\key{Sort} 0, \key{Sort} 1)\#i} etc., leading to an infinite typing derivation.
\end{example}
\autoref{ex:lea} and \autoref{sec:tensor} demonstrate the interplay between tuples/packs/$\Sigma$-types/arrays, their normalizations, and dependent types.

\mimir also provides an \hl{\key{insert} (e$_t$, e$_i$, e$_v$)} operation, that we elided in the formal presentation.
It non-destructively creates a new tuple where the element at index~\hl{e$_i$} has been replaced with~\hl{e$_v$}.
From a semantic point of view, insertion is a mix of term elimination and introduction as, for instance, \hl{\key{insert} (e$_t$, 1$_3$, e$_v$)} is the same as \hl{(e$_t$\#0$_3$, e$_v$, e$_t$\#2$_3$)}.

\subsubsection{Assignable}
\label{sec:assignable}

Consider \hl{(\key{Nat}, \%core.ncmp.l)} of type \hl{[*, [\key{Nat}, \key{Nat}] $\to$ \key{Bool}]}.
However, we can also type it as
\begin{lstlisting}
    let Cmp = [T: *, [T, T] -> Bool]
\end{lstlisting}
which is similar to a trait in Rust or type class in Haskell.
Most systems featuring existential or $\Sigma$-types require tuples to be \emph{ascribed} such as \hl{(\key{Nat}, \%core.ncmp.l):Cmp}.
In \mimir, tuples are \emph{not} ascribed.
Instead, whenever an expression \hl{e} is \emph{assigned} to a variable \hl{x:T}, \mimir checks via $\atype{e}{T}$ whether this assignment actually makes sense.
This is trivially the case, if \hl{e}'s type is \hl{T} (\iref{A-T}).
Otherwise, \iref{A-Tup} recursively checks whether all elements of a tuple are assignable while successively resolving the dependencies the $\Sigma$-type~\hl{T} may introduce (similar to \irefx{Ex-SL}{Ex-S$_L$}/\irefx{Ex-SI}{Ex-S$_i$} discussed above).
\begin{example}
    In the following code the type checker allows passing \hl{(\key{Nat}, \%core.ncmp.l)} to \hl{f}, although it expects an instance of \hl{Cmp}:
    \irefrule{Lam} asks \irefrule{A-Tup} whether the given pair is assignable to \hl{Cmp}, which it is in fact.
    \begin{lstlisting}
        lam f(T: *, less: [T, T] -> Bool)(x: T): Bool = less (x, x);
        f (Nat, %core.ncmp.l) 23
    \end{lstlisting}
\end{example}

\subsubsection{Functions}\label{sec:fun}

A function \hl{$\lambda$x:T@e$_f$:U = e} has a dependent function type \hl{[x:T] $\to$ U} (\iref{Lam})---written as \hl{$\Pi$ x:T.U} in literature.
This means that the \emph{type} of the function's codomain \hl{U} may depend on the \emph{value} of the argument~\hl{x} which in turn must inhabit the domain~\hl{T}.
%A $\Pi$-type either lives in \lstinline|*|, if both its domain and codomain type as \lstinline|*|, or in $\square$, if at least one of them types as $\square$ (\iref{Pi}).
%As usual \lstinline|T $\to$ U| is shorthand for \lstinline|PI_: T $\to$ U|.
The callee \hl{e} of an application \hl{e e$_T$} must type as a $\Pi$-type and the given argument~\hl{e$_T$} must be \emph{assignable} (see above) to the domain;
the type of the application is resolved by substituting \hl{x} with the argument in the codomain~\hl{U}~(\iref{App}).
The Boolean term \hl{@e$_f$} is the so-called \emph{filter} and is used for partial evaluation (\autoref{sec:sema_pe}).

\paragraph{\Ac{CPS}}

Continuations are functions that never return.
\mimir models them as functions whose codomain is \hl{$\bot$}~(\iref{Bot})---a type without inhabitants---and \mim provides syntactic sugar (\autoref{fig:simple_sugar}) for continuations (\lstinline|cn|/\lstinline|con|) and their types (\lstinline|Cn|).
Continuations bridge the gap between \acp{CFG} and the $\lambda$-calculus, as continuations are akin to basic blocks.
\begin{example}\label{ex:phi}
    Consider the \enquote{if diamond} in \autoref{fig:cond_text}.
    The expression \hl{(F, T)\#cond ()} first selects either the \hl{F} or \hl{T} continuation.
    This works because \hl{cond} is of type \lstinline|Bool|---an alias for \mbox{\hl{\key{Idx} 2}}.
    The result of the extraction is of type \hl{\key{Cn} []} (see \irefx{Ex-SI}{Ex-S$_i$} and \irefx{N-PackB}{N-Pack$_\beta$}) which allows us to apply the selected continuation to~\hl{()} to continue execution there.
    The \hl{T} case invokes \mbox{\hl{N 42}} while the \hl{F} case invokes \hl{N 23}.
    In \ac{SSA} form, \hl{f}, \hl{F}, \hl{T}, and \hl{N} would be basic blocks and \hl{N} would need a $\phi$-function to select either \hl{23} or \hl{42}.
    In \ac{CPS} the $\phi$-function becomes the parameter \hl{phi} of continuation \hl{N} while the operands of the $\phi$-function become arguments to the appropriate continuation call~\cite{DBLP:journals/sigplan/Appel88,DBLP:conf/irep/Kelsey95}.
    See also \autoref{ex:if}.
\end{example}
Continuations may be used (mutually) recursive (\autoref{sec:rec}) to model arbitrary, unstructured control flow.
Note that \hl{f} in \autoref{fig:cond_text} is a continuation and the \hl{return} point is explicit by passing it to \hl{f} as another continuation.
This idiom is so common that \mim introduces \lstinline|fn|/\lstinline|fun|/\lstinline|Fn| as sugar (\autoref{fig:simple_sugar}).
For example, in \autoref{fig:cond_text} we can instead write:
\begin{lstlisting}
fun f(cond: Bool): Nat = /*...*/;
\end{lstlisting}
Finally, \mim provides syntactic sugar for \emph{curried} functions \& continuations (\autoref{lst:fun_expr}-\subref{lst:fun_type}).
Note that all curried function groups except the last one are in direct style in the case of curried continuations.

\subsection{Full Recursion}
\label{sec:rec}

Named functions (\autoref{fig:simple_sugar}) are in fact more powerful than ordinary \lstinline|let| bindings, as they allow for (mutual) recursion and, hence, are more like \verb|letrec| in other languages.
\begin{lstlisting}
    lam forever(x: Nat): Nat = forever x;
\end{lstlisting}
This is an extension of the calculus presented so far.
Internally, the recursive function \hl{forever} is represented as a cyclic graph (\autoref{sec:graph}).
\mimir allows recursion in both direct style (\autoref{ex:pow}) and \ac{CPS}:
%\mimir supports \ac{CPS} to allow for complex control flow.
\begin{example}\label{ex:count}
    The following \hl{loop} counts from \hl{0} to \hl{42}.
    If its parameter \hl{i} (the \enquote{$\phi$-function}) is less than \hl{42}, \hl{i} is incremented and \hl{loop} recurses.
    Otherwise, it \mbox{\hl{exit}s}.
    \begin{lstlisting}[mathescape=true,literate=]
        con loop(i: Nat) =                                // i = $\color{codegreen}\phi$(0, i + 1)
            (exit, body)#(%core.ncmp.l (i, 42)) () where  // if i < 42 then body() else exit()
                con body() = loop (%core.nat.add (i, 1)); // recurse
                con exit() = next i;                      // pass last instance of i = 42 to next
            end;
        loop 0;                                           // loop entry
    \end{lstlisting}
\end{example}

\subsection[Partial Evaluation \& beta-Equivalence]{Partial Evaluation \& $\beta$-Equivalence}
\label{sec:sema_pe}

The so-called \emph{filter} \hl{@e$_f$} of a function is used for partial evaluation and is a Boolean expression that may depend on the function's parameter (\irefx{N-B}{N-$\beta$}).
For each call-site, \mimir instantiates the filter by substituting the function's variable with the call-site's argument.
Remember that substitution also normalizes.
If this syntactically yields \lstinline|tt|, \mimir will $\beta$-reduce this call-site---potentially recursively inlining more function calls.
This mechanism allows for compile-time specialization. % and the first \citeauthor{Futamura:1999:PEC:609149.609205} projection enables you to write interpreters instead of compilers (\autoref{sec:overview}).

Since \mimir has a dependent type system featuring full recursion (\autoref{sec:rec}), type-checking becomes undecidable in general since checking for $\beta$-equivalence is.
This is arguably the most concerning issue for picking up dependent types in more mainstream programming languages.
\mimir tackles this issue by (partially) evaluating functions according to the filters during normalization.
\begin{example}\label{ex:pow}
    The following recursive function \hl{pow}, will be recursively $\beta$-reduced, if the exponent \hl{b} is a compile-time constant.
    Note that \hl{f}'s parameter \hl{x} has a dependent type.
    \begin{lstlisting}
        lam pow(a b: Nat)@%core.pe.known b: Nat =
            (%core.nat.mul (a, pow(a, %core.nat.sub (b, 1))), 1)#(%core.ncmp.e (b, 0));

        lam f(n: Nat, x: <<%core.nat.mul (n, %core.nat.mul (n, n)); Nat>>): [] = ();
        lam g(m: Nat, y: <<pow (m, 3); Nat>>): [] = f (m, y);
    \end{lstlisting}
    As \hl{g} calls \hl{f} with \hl{(m, y)}, \mimir has to check whether \hl{y}'s type is assignable to \hl{g}'s domain (\iref{Pi}).
    \irefrulex{N-B}{N-$\beta$} determines that \hl{pow}'s filter evaluates to \lstinline|tt| with the given argument \hl{(m, 3)}.
    This causes \mimir to immediately $\beta$-reduce \hl{pow (m, 3)} which will result in a call-site \hl{pow (m, 2)}.
    \irefrulex{N-B}{N-$\beta$} determines again that the filter yields \lstinline|tt|, causing another $\beta$-reduction.
    The type checker now sees \hl{y}'s type as
    \begin{lstlisting}
        <<%core.nat.mul (m, %core.nat.mul (m, m)); Nat>>
    \end{lstlisting}
    which is assignable to \hl{g}'s domain.
    Using a \lstinline|ff| filter for \hl{pow} would result in a type error, as \hl{x}'s type is different from \hl{<<pow (m, 3); \key{Nat}>>}.
    Using a \lstinline|tt| filter for \hl{pow} is dangerous:
    While the example above would still work, a call-site like \hl{pow (i, j)}, where \hl{j} is not a compile-time constant, would cause \mimir to diverge, as it would endlessly $\beta$-reduce new calls to \hl{pow}.
    However, this termination behavior is not random and completely transparent to the programmer.
    It depends on the specified filters as well as how they are used---\emph{as stated by the programmer}.
\end{example}
Coupling partial evaluation with normalization that in turn interacts with type checking is a novel approach to dependent type checking.
Moreover, partial evaluation filters clearly determine which parts of a program are evaluated at compile time and which ones remain in the compiled program.
This distinction is somewhat unclear in many dependently typed languages---Idris~2 is an exception (see \autoref{sec:relwork}).

Note that the rules in \autoref{fig:rules} do \emph{not} contain a conversion rule that states that two terms are equal modulo $\beta$-equivalence.
The typing rules are syntax-directed and deterministic similar to \citet{pollack92}.
The normalization rules unambiguously state where normalization and, in particular, $\beta$-reduction happens during type checking.

\subsubsection{Default Filter Policy}
\label{sec:filter}

Since our goal is high-performance code, we want to specialize type variables at compile time~\cite{DBLP:journals/jfp/TolmachO98}---similar to C++ templates, Rust generics, polymorphism in MLton~\cite{DBLP:journals/jfp/ShivkumarMZ21} but unlike Java generics.
For this reason, \mim defaults to \lstinline|tt| for elided partial evaluation filters except for the final continuation in a (curried) \lstinline|cn|/\lstinline|con|/\lstinline|fn|/\lstinline|fun| function.
The filter of this final continuation defaults to \lstinline|ff| (\autoref{fig:sugar}).
This has the desired effect that the actual computations are deferred to runtime while other abstractions such as type abstractions are specialized.
However, programmers can override the default behavior by explicitly specifying a filter as in \autoref{ex:pow}.

\subsection{Type Safety}
\label{sec:meta}

In order to argue about \mimir's soundness, we present a nondeterministic $\beta$-reduction relation $\step{e}{e}$ that \emph{steps} a normalized expression---possibly by descending into a subexpression via \iref{Cong}---into another normalized expression by performing a single $\beta$-reduction~(\iref{Beta}).
\irefrule{Cong} introduces nondeterminism as the evaluation context~$\mathcal{E}$ comprises all possible evaluations for expressions.
This provides the flexibility during optimization, for instance, to either descend into the callee or the argument of an application---or directly apply \iref{Beta}, if possible.

We have formalized the rule system in the proof assistant Coq and the following lemmas establish \mimir's type safety (recall once again that all expressions are normalized):
\begin{lemma}[Progress]\label{lem:progress}
    If $\hlgamma \vdash \hl{e} : \hl{T}$, then \hl{e} is a value or there exists an \hl{e'} such that $\step{e}{e'}$.
\end{lemma}
\begin{lemma}[Preservation]\label{lem:preservation}
    If $\hlgamma \vdash_\beta \hl{e} : \hl{T}$,  and $\step{e}{e'}$, then $\hlgamma \vdash_\beta \hl{e'} : \hl{T}$.
\end{lemma}
Since dependent types may depend on variables introduced by lambdas, a \iref{Beta}-step may change the type of an expression.
For this reason, \mimir does not guarantee \emph{strong} preservation, as the resulting type after a step may differ from the original one.
However, the preservation is stronger than \emph{weak} preservation since the type of the expression after a step is not arbitrary but evolves from the old one by zero or more $\to$-steps.
To model this, we introduce $\beta$-equivalence on type level as indicated in the typing rule~$\vdash_\beta$.
This is a technical addition to prove preservation as seen in other models of \ac{CC}.

For a proper evaluation semantics you can choose any deterministic subset of $\to$ such as a strict, left-to-right evaluation order as in \mimir's \llvm backend.
However, the nondeterministic relation~$\to$ also establishes type safety for reductions that happen a~priori.
For example, many optimizations such as copy propagation or scalarization are implemented on top of $\beta$-reductions.

\subsubsection{Composition of Plugins}

Custom axioms/normalizers, however, could violate progress or preservation.
Violation of preservation is a bug in the plugin that will trigger a type error as soon as the erroneous normalization fires.
For example, the nonsensical normalization $\hl{3 + 5} \vtr \hl{()}$ changes the type from \lstinline|Nat| to \hl{[]}.
\mimir will catch such errors after applying a non-type-preserving rewrite, as this will lead to an ill-typed program.
Specifying rewrite rules directly in \mim (see end of \autoref{sec:norm}) would allow us to directly type-check rules for preservation even before firing them.

In addition, an axiom may violate progress, if the function it represents is not total.
This means that the axiom's normalizer (or the translation of the axiom into another program) cannot cope with all possible inputs.
For example, \mbox{\hl{\%mem.load}ing} from a dangling pointer or \mbox{\hl{\%core.bitcast}ing} $\hl{4$_5$}$ to $\hl{\key{Idx} 4}$ are such unsafe instances and lead to undefined behavior.
Violating progress is not necessarily bad per se.
For example, if you want to compile an unsafe language like C, you obviously need an unsafe language to model this behavior.
If you want your axioms to be type safe, you must guarantee that they do not violate progress.

Even though axioms do not provide an implementation by themselves, they are treated and, in particular, type-checked like any other value in \mimir.
As long as the axioms' implementations obey preservation and progress, Lemma~\ref{lem:progress} and~\ref{lem:preservation} guarantee a type-safe composition which in turn enables a type-safe composition of plugins.
Even if an axiom violates progress, it is merely a problem inherent to that axiom, and does not arise from interactions with other axioms.
In addition, axioms are in contrast to instructions in \mlir, \llvm, or \thorin \emph{first-class} citizens.
This means, for instance, that an axiom or a partial application of it can be passed as an argument to another function or higher-order axiom.
See \hl{\%tensor.reduce} in \autoref{sec:tensor} for an example and \autoref{sec:eval:discussion} for our practical experience with composing plugins.

\subsubsection{Type-Safe Code Transformations}\label{sec:loop-unroll}

Oftentimes, code transformations are expressible as specializations.
For example, suppose we want to create several unrolled loops---each time with a different body and differently typed induction variables.
\mim allows us to directly specify a polymorphic loop and partially evaluate it to generate the unrolled, specialized version.
The following function \lstinline|iter| invokes \lstinline|n| times \lstinline|body| and jumps to \lstinline|exit| afterwards.
The parameter \lstinline|body| expects another continuation as argument which is the \lstinline|continue| point within \lstinline|iter|.
In order to allow for arbitrary induction variables, \lstinline|iter| is polymorphic in its accumulator type \lstinline|A| that the other continuations use to communicate between different iterations.
\begin{lstlisting}
con iter {A: *} (n: Nat) (body: Cn [Cn A][A]) (exit: Cn A) (acc: A)@tt =
    head(0, acc) where
        con head (i: Nat, acc: A)@%core.pe.known n =
            let cond = %core.ncmp.l (i, n);
            (f, t)#cond () where
                con t() = body continue acc where
                    con continue(acc: A) = head ((%core.nat.add (i, 1)), acc);
                end
                con f() = exit acc;
            end;
    end;
\end{lstlisting}
Now, if we invoke
\begin{lstlisting}
    iter n body exit (0, 1)
    where
        con body (yield: Cn [Nat, Nat]) (a b: Nat) = yield (b, %core.nat.add (a, b));
        con exit (a _: Nat) = return a;
    end
\end{lstlisting}
we essentially construct a \lstinline|while| loop that uses in addition to the loop counter \lstinline|i| two \lstinline|Nat|s as accumulator (the loop's \enquote{$\phi$}-functions---see \autoref{ex:phi}) and computes the \lstinline|n|th Fibonacci number.
If we change the call to use \lstinline|12| instead of \lstinline|n|, the filter of \lstinline|head| recursively becomes \lstinline|tt| which causes a complete unrolling of the loop and results in a cascade of~12 continuations.
This will also happen, if calling \lstinline|iter| from C++ (see \autoref{sec:cpp}).
After a standard set of optimizations the result is just \lstinline|return 144|---the 12th Fibonacci number.

Note that in \acp{IR} like \mlir or \llvm we would write C++ code, instead, that \emph{generates} an unrolled loop specialized for the desired type (two \lstinline|Nat|s in our example).
This C++ "generator"-code is significantly more complex because writing code that emits code is inherently more verbose.
In addition, we have to \enquote{look through} the API calls and picture in our minds how the generated program will look like.
Furthermore, this generator code is also significantly more error-prone because we may very well accidentally construct ill-typed \ac{IR}.
This cannot happen, if our starting point is already well-typed \mim as above.
% Finally, working with programs that generate programs that generate programs (TableGen $\to$ C++ $\to$ \mlir) is cumbersome to work with:
% The build process becomes more and more complex and it becomes increasingly more unclear how exactly a snippet of \mlir code was conjured up.

\FloatBarrier
% vim:spell:spelllang=en
\newcommand{\myarrow}[1][]{\mathrel{\tikz{
    \node[fill, minimum width=.5ex, minimum height=.9em, inner sep=0pt, single arrow, single arrow head extend=2pt, single arrow tip angle=60]
            (A){\raisebox{3.5pt}[0pt][0pt]{$\,\scriptstyle #1\ $}}; \path([xshift=-.4pt]A.west)--(A.east);}}}

\section{Graph Representation}
\label{sec:graph}

\begin{figure}[t]
    \centering
    \begin{subcaptionblock}{.31\textwidth}
        \begin{lstlisting}[basicstyle=\ttfamily\tiny,xleftmargin=1ex,xrightmargin=1ex]
            let a = %core.wrap.add 0 (x,1I8);
            let b = %core.wrap.add 0 (x,2I8);
        \end{lstlisting}
        \vspace{-1.5ex}
        \caption{Two additions in \mim}
        \label{fig:add_text}
        \vspace{.5ex}
        \begin{lstlisting}[basicstyle=\ttfamily\tiny,xleftmargin=1ex,xrightmargin=1ex]
            con f(cond: Bool, ret: Cn Nat) =
              (F,T)#cond () where
                con F() = N 23;
                con T() = N 42;
                con N(phi: Nat) = ret phi;
              end;
        \end{lstlisting}
        \vspace{-1.ex}
        \caption{\enquote{If diamond} w/ \enquote{phi} in \mim}
        \label{fig:cond_text}
    \end{subcaptionblock}
    \begin{subcaptionblock}{.33\textwidth}
        \centering
        \begin{tikzpicture}[
            node distance=.3cm and .25cm,
            n/.style={rectangle,draw=ACMRed,      fill=ACMOrange,    rounded corners, text centered},
            s/.style={rectangle,draw=ACMDarkBlue, fill=ACMLightBlue, rounded corners, text centered},
        ]
            \node[s](n32)                         {\lst[basicstyle=\ttfamily\tiny]|256|};
            \node[s](n0)   [right=of n32]         {\lst[basicstyle=\ttfamily\tiny]|0|};
            \node[s](add)  [left=of n32]          {\lst[basicstyle=\ttfamily\tiny]|%core.wrap.add|};
            \node[s](app1) [below=of add]   {\lst[basicstyle=\ttfamily\tiny]|App|};
            \node[s](app2) [below right=of app1]  {\lst[basicstyle=\ttfamily\tiny]|App|};
            \node[s](app3) [below right=of app2]  {\lst[basicstyle=\ttfamily\tiny]|App|};
            \node[s](app4) [right=of app3]        {\lst[basicstyle=\ttfamily\tiny]|App|};
            \node[s](x1)   [above=of app3]        {\lst[basicstyle=\ttfamily\tiny]|Tuple|};
            \node[s](x2)   [above=of app4]        {\lst[basicstyle=\ttfamily\tiny]|Tuple|};
            \node[s](x)    [above=of x1]          {\lst[basicstyle=\ttfamily\tiny]|x|};
            \node[s](n1)   [above right=of x1]    {\lst[basicstyle=\ttfamily\tiny]|1|};
            \node[s](n2)   [above right=of x2]    {\lst[basicstyle=\ttfamily\tiny]|2|};
            \draw [-latex] (app1) -- (add);
            \draw [-latex] (app1) -- (n32);
            \draw [-latex] (app2) -- (app1);
            \draw [-latex] (app2) to[bend left=10] (n0);
            \draw [-latex] (app3) -- (app2);
            \draw [-latex] (app3) -- (x1);
            \draw [-latex] (app4) -- (app2);
            \draw [-latex] (app4) -- (x2);
            \draw [-latex] (x1) -- (x);
            \draw [-latex] (x1) -- (n1);
            \draw [-latex] (x2) -- (x);
            \draw [-latex] (x2) -- (n2);
        \end{tikzpicture}
        \caption{\mimir graph for \ref{fig:add_text}}
        \label{fig:add_graph}
    \end{subcaptionblock}
    \begin{subcaptionblock}{.34\textwidth}
        \centering
        \begin{tikzpicture}[
            node distance=.3cm and .5cm,
            n/.style={rectangle,draw=ACMRed,      fill=ACMOrange,    rounded corners, text centered},
            s/.style={rectangle,draw=ACMDarkBlue, fill=ACMLightBlue, rounded corners, text centered},
        ]
            \node[n] (main)                 {\lst[basicstyle=\ttfamily\tiny]|con f(cond: Bool, return: Cn Nat)|};
            \node[s] (cond) [below=of main] {\lst[basicstyle=\ttfamily\tiny]|(F, T)#cond ()|};
            \node[n] (F)    [ left=of cond] {\lst[basicstyle=\ttfamily\tiny]|con F()|};
            \node[n] (T)    [right=of cond] {\lst[basicstyle=\ttfamily\tiny]|con T()|};
            \node[s] (FF)   [below=of F]    {\lst[basicstyle=\ttfamily\tiny]|N 23|};
            \node[s] (TT)   [below=of T]    {\lst[basicstyle=\ttfamily\tiny]|N 42|};
            \node[n] (N)    [ right=of FF]  {\lst[basicstyle=\ttfamily\tiny]|con N(phi: Nat)|};
            \node[s] (NN)   [ below=of N]   {\lst[basicstyle=\ttfamily\tiny]|return phi|};
            \draw [-latex] (main) -- (cond);
            \draw [-latex] (cond) -- (F);
            \draw [-latex] (cond) -- (T);
            \draw [-latex] (F) -- (FF);
            \draw [-latex] (T) -- (TT);
            \draw [-latex] (FF) -- (N);
            \draw [-latex] (TT) -- (N);
            \draw [-latex] (N) -- (NN);
            %\draw [-latex] (cond) to[bend right=25] (main);
            \draw [-latex] ($(cond)+(0.1cm,+0.1cm)$) -- ($(main)+(-0.8cm,-0.05cm)$);
            %\draw [-latex] (TT) to[bend right=90] (main);
            %\draw [-latex] (FF) to[bend  left=90] (main);
            %\draw [-latex] (NN) to[xshift=1cm,yshift=.5cm] (N);
            \draw [-latex] ($(NN)+(0.4cm,+0.1cm)$) -- ($(N)+(-0.0cm,-0.05cm)$);
            \draw [-latex] ($(NN)+(-0.2cm,+0.1cm)$) to[out=150,in=-80] ($(main)+(0.5cm,-0.1cm)$);
        \end{tikzpicture}%
        %\vspace{-4ex}
        \caption{Condensed \mimir graph for \ref{fig:cond_text}}
        \label{fig:cond_graph}
    \end{subcaptionblock}
    \vspace{-2ex}
    \caption{\mim vs.~\mimir. Type edges are elided. Immutables are in blue, mutables in orange.}
\end{figure}

\mimir's implementation is based upon the \enquote{sea of nodes} concept~\cite{DBLP:conf/irep/ClickP95}.
This means that \mimir's internal representation is a data dependence graph where each node in the graph represents an expression.
If \hl{e$_s$} is a subexpression/operand of an expression~\hl{e}, the graph contains an edge~$\hl{e} \myarrow \hl{e$_s$}$.
%Which expressions need which subexpressions is directly apparent from the syntax (\autoref{fig:syntax}).
For example, an \emph{App} has two operands---the callee and argument---and, hence, two outgoing edges while a $\Sigma$-type with three element types has three operands and, hence, three outgoing edges.
In addition, every expression~\hl{e} has exactly one type~\hl{T} it inhabits.
This relation is modeled with another edge~$\hl{e} \rightarrowtriangle \hl{T}$ in the program graph.
Note once again that \hl{T} is an expression.
Hence, there is only \emph{one} graph that contains both terms and types as well as any dependencies between them via edges;
in particular, we may have types that depend on terms due to dependent types.
Furthermore, this graph is \emph{complete}.
This means that this graph comprises the whole semantics of a \mimir program and---apart from an internal hash set to hash-cons all nodes (\autoref{sec:overview:norm})---\mimir does \emph{not} rely on any other auxiliary data structures like instruction lists, basic blocks, special regions, or \acp{CFG}.
\lstinline|let|-expressions and even explicit function nesting only exist in \mim.
Thus, a \mimir graph \emph{solely} consists of a large number of nodes that \enquote{float} in a complex network, resembling a sea---hence the term \enquote{sea of nodes}.

\begin{wraptable}{r}{.45\textwidth}
    \centering
    \vspace{-3ex}
    \begin{footnotesize}%
    \begin{tabular}{ll}
        \toprule
        Immutable & Mutable \\
        \midrule
        %\emph{must be} \lst[language=C++]|const| & \emph{may be} non-\lst[language=C++]|const| \\
        build operands first, & build the node first, \\
        \ \ then the actual node & \ \ then set the operands \\
        operands form a DAG & operands may be cyclic \\
        hash-consed & each new entity is fresh \\
        non-parametric & may be parametric \\
        %\texttt{rebuild} & \texttt{stub} \\
        \bottomrule
    \end{tabular}
    \end{footnotesize}%
    \caption{Immutable vs.~mutable nodes}
    \vspace{-3ex}
    \label{tab:mut}
\end{wraptable}

\begin{example}
    \autoref{fig:add_graph} depicts the \mimir graph of the \mim program in \autoref{fig:add_text}.
    Note that the \lstinline|let|-expressions in \mim are absent in \mimir and the curried call \lstinline|%core.wrap.add /*256*/ 0|, where the inferred implicit argument is now explicit, is shared by both additions.
    In fact, all 8-bit wide additions with mode \hl{0} (\autoref{sec:core_plugin}) will reuse this subgraph.
    Similarly, \hl{x} is shared by both argument tuples of the additions.
    What is not shown in \autoref{fig:add_graph} are the type edges.
    For example, the bottom Apps and the literals \hl{1} and \hl{2} point to a subgraph that constitutes \mhl{\key{Idx} 256}.
\end{example}

\paragraph{Immutables vs. Mutables}
%\label{sec:mutable}

So far, we have discussed expressions that we create by first building their operands, and then the actual nodes.
\mimir calls these expressions \emph{immutable}:
Once constructed, they cannot be changed later on.
Hence, immutables form a \ac{DAG}.
In order to allow for recursion, we have to somehow form a cyclic graph.
For this reason, there are also \emph{mutable} expressions (\autoref{tab:mut}).
For mutables, we first build the node and set its operands later on.
This enables us to form a cyclic graph and therefore enables recursion.
Mutables also allow changing their operands later on---hence their name.
This also means that a mutable will not be hash-consed as opposed to immutables.
But \mimir will check mutables for $\alpha$-equivalence, if necessary.
Finally, all expressions (both on term and type level) that introduce variables are also modeled as mutables.

\begin{example}\label{ex:if}
    Consider the \mimir graph in \autoref{fig:cond_graph} of \autoref{fig:cond_text}.
    Note how this graph resembles a classic \ac{CFG}.
    In order to access \hl{f}'s and \hl{N}'s variables, these continuations are mutables.
    In fact, most of the time we build functions as mutables even though \hl{F} and \hl{T} could be modeled as immutables in this particular case, as they are neither used recursively nor are their variables accessed.
\end{example}

% vim:spell:spelllang=en
\section{Type Checking, Inference, and Normalization On-The-Fly}
\label{sec:impl}

\mimir's implementation performs normalization, type checking, and inference of implicits as soon as a new expression is constructed.
This eagerness has the advantage that a compiler/plugin developer will immediately notice, if they incorrectly plugged together some expressions that are ill-typed.

\subsection[alpha-Equivalence]{$\alpha$-Equivalence}

This means that type checking, inference, and normalization must also work on open terms, i.e., in the presence of free variables.
%When checking for $\alpha$-equivalence, \mimir also simultaneously try to resolve implicits.
The implementation boils down to checking for $\alpha$-equivalence modulo free variables inside of the \emph{assignable} relation.
Here we know that \emph{either} two expressions \emph{must} be $\alpha$-equivalent \emph{or} there is a type error.
For this reason, \mimir optimistically assumes during \emph{type checking} that any free variables are $\alpha$-equivalent.
Eventually, all terms will be closed where any remaining issues will be found.
\begin{example}
    During type checking \mimir considers the following expressions as $\alpha$-equivalent:
    \begin{equation*}
        \lstinline|LAM(a: Nat): Nat = b| \qqtext{and} \lstinline|LAM(x: Nat): Nat = y|\ .
    \end{equation*}
\end{example}
\begin{example}
    For \emph{normalization}, however, this assumption is unsound.
    \iref{N-PackTup} cannot simply normalize the following expression as \lstinline|b| and \lstinline|y| may be bound differently:
    \begin{equation*}
        \lstinline|(LAM(a: Nat): Nat = b, LAM(x: Nat): Nat = y)| \ntriangleright \lstinline|<2; LAM(a: Nat): Nat = b>|
    \end{equation*}
\end{example}

As discussed in \autoref{sec:graph}, pointer equality of two \mimir expressions implies normalized, syntactic equivalence.
Both expressions refer to the same object.
However, pointer equality does not necessarily imply $\alpha$-equivalence when free variables are involved.
For this reason, \mimir only resorts to pointer equality when checking for $\alpha$-equivalence in the absence of free variables.
\begin{example}
    Although the bodies of the following functions enjoy pointer equality, these functions are \emph{not} $\alpha$-equivalent as \lstinline|x| is bound in the first function and free in the second one:
    \begin{equation*}
        \lstinline|LAM(x: Nat): Nat = x| \qqtext{and} \lstinline|LAM(y: Nat): Nat = x|
    \end{equation*}
\end{example}

\subsection{Type Inference}
\label{sec:infer}

Whenever a function with an implicit argument is invoked, the function will first be applied with a placeholder.
\begin{example}
    \label{ex:minus}
    Consider the annex \lstinline|%core.minus| that computes unary minus of its \lstinline|s|-sized \lstinline|Idx| argument \lstinline|a| (\autoref{lst:core}).
    Note that \lstinline|s| is implicit.
    Furthermore, the operation expects a so-called mode of type \lstinline|Nat| (see \autoref{sec:case:ir} for details).
    As soon as we apply the first explicit argument \lstinline|mode| to \lstinline|%core.minus|, \mimir will insert a fresh placeholder---let us say \lstinline|?5|---as implicit argument in between:
    \begin{lstlisting}
        %core.minus ?5 mode
    \end{lstlisting}
    This yields the type \lstinline|Idx ?5 -> Idx ?5| (see \iref{App}).
    When we apply \lstinline|42$_{256}$| as third argument, \iref{App} triggers the \emph{assignable} relation.
    Rules \iref{A-T} and \iref{A-Tup} additionally match placeholders with the provided argument and fill out any gaps.
    So here, we find: \lstinline|?5| = \lstinline|256|:
    \begin{lstlisting}
        %core.minus /*256*/ mode 42$_{256}$
    \end{lstlisting}
    Due to the implicit \lstinline|tt| filter, \mimir will $\beta$-reduce the application to:
    \begin{lstlisting}
        %core.wrap.sub /*256*/ mode (%core.idx 256 mode 0, 42$_{256}$)
    \end{lstlisting}
    This in turn will be normalized to \lstinline|214$_{256}$| (two's complement of -42).
\end{example}

\subsection{C++ Interface}
\label{sec:cpp}

Type checking, inference, and normalization happens \emph{regardless} of whether \mimir is used through \mim or its C++ interface.
% In fact, the \mimir parser directly emits a \mimir graph.
\begin{example}
    To better bring the point across how \mimir behaves when controlling from C++, reconsider \autoref{ex:minus}.
    We can construct the same call using the C++-API:
    \begin{lstlisting}[language=C++,morekeywords=int8_t,literate=]
        const Def* res = world.call<core::minus>(mode, world.lit_idx(256, 42));
    \end{lstlisting}
    Now, type checking, inference, and normalization will happen as discussed above.
    Thus, the C++ variable \lstinline|res| (of static type \lst[language=c++]|const Def*| and dynamic type \lst[language=c++]|const Lit*|) will point to a literal that represents \lstinline|244| of type \lstinline|Idx 256|.
    In other words, creating \mimir expressions from C++ triggers the usual normalization that may in turn cause a Turing-complete (partial) evaluation of the expression.
\end{example}
\begin{example}
    When trying to construct an ill-typed \mimir program from C++, \mimir will throw a C++ exception as in the following code that erroneously passes a \lstinline|Nat| instead of an \lstinline|Idx| literal to \lstinline|%core.minus|:
    \begin{lstlisting}[language=C++,morekeywords=int8_t,literate=]
        const Def* res = world.call<core::minus>(mode, world.lit_nat(42)); // %core.minus mode 42
    \end{lstlisting}
\end{example}

% vim:spell:spelllang=en
\section{Case Studies \& Evaluation}
\label{sec:eval}

Our case studies show dependent types as a useful tool in regards to safe guards, expressiveness, and efficiency.
Informative types prevent unwanted behavior and allow for more aggressive optimizations.
This section showcases \mimir's flexibility by discussing some of the plugins we have already developed and evaluating their performance.
In particular, we want to show that
    \mimir is able to achieve the same performance as other low-level approaches like C/\llvm,
    is able to fully remove carefully crafted abstractions, and that
    designing code analyses/transformations in \mimir is no more complicated than in traditional compiler \acp{IR} like \llvm.

If not mentioned otherwise, we ran all tests using a single thread on an AMD Ryzen 7 3700X supported by 64GB of DDR4@2133MT/s RAM.
We used \llvm/Clang 15.0.7 for the C sources as well as the \llvm code emitted by \mimir.
All \mimir programs compiled within a few seconds at most.

% vim:spell:spelllang=en
\subsection{Low-Level Plugins}
\label{sec:case:ir}

In this section, we discuss the design of three low-level plugins, which are deliberately modeled on \llvm:
First of all, \llvm is a well-thought-out low-level \ac{IR};
second, it allows for straightforward mapping of axioms defined in these plugins to \llvm instructions in \mimir's \llvm backend.
This backend also lives within a plugin.
Other more high-level plugins eventually lower their axioms to those of these low-level plugins.
The \verb|%core| plugin introduces integer operations,
the \verb|%math| plugin a floating-point type operator and operations for it, and
the \verb|%mem| plugin side-effects, memory operations, and pointer arithmetic.

\begin{figure}[t]
    \begin{subcaptionblock}{\textwidth}
        \begin{lstlisting}
            // Create literal of type Idx s from l while obeying mode m
            axm %core.idx: [s: Nat] [m: Nat] [l: Nat] -> Idx s;

            axm %core.nat(add, sub, mul): [a b: Nat] -> Nat;
            axm %core.ncmp(/*...*/):      [a b: Nat] -> Bool;

            axm %core.bit1(f, neg, id, t):      {s: Nat} [m: Nat] [a  : Idx s] -> Idx s;
            axm %core.bit2(/*...*/):            {s: Nat} [m: Nat] [a b: Idx s] -> Idx s;
            axm %core.wrap(add, sub, mul, shl): {s: Nat} [m: Nat] [a b: Idx s] -> Idx s;
            axm %core.shr(a, l):                {s: Nat}          [a b: Idx s] -> Idx s;
            axm %core.icmp(/*...*/):            {s: Nat}          [a b: Idx s] -> Bool;

            axm %core.conv(s, u): {ss: Nat} [ds: Nat] [Idx ss] -> Idx ds;
            axm %core.bitcast:   {S: *} [D: *] [S] -> D;

            lam %core.minus {s: Nat} (m: Nat) (a: Idx s): Idx s = %core.wrap.sub m (%core.idx s m 0, a);
        \end{lstlisting}
        \vspace{-1.ex}
        \caption{The \texttt{\%core} plugin}
        \vspace{1.ex}
        \label{lst:core}
    \end{subcaptionblock}
    \begin{subcaptionblock}{\textwidth}
        \begin{lstlisting}
            axm %math.F: [p e: Nat] -> *;
            let %math.f64 = (52, 11);          // similar: f16, f32, ...
            let %math.F64 = %math.F %math.f64; //          F16, F32, ...

            axm %math.arith(add, sub, mul, div, rem):
                                    {p e: Nat} [m: Nat] [a b: %math.F (p, e)] -> %math.F (p, e);
            axm %math.tri(/*...*/): {p e: Nat} [m: Nat] [a  : %math.F (p, e)] -> %math.F (p, e);
            axm %math.cmp(/*...*/): {p e: Nat} [m: Nat] [a b: %math.F (p, e)] -> Bool;

            lam %math.minus .(p e: Nat) (m: Nat) (a: %math.F (p, e)): %math.F (p, e) =
                %math.arith.sub m (0:(%math.F pe), a);
        \end{lstlisting}
        \vspace{-1.ex}
        \caption{The \texttt{\%math} plugin}
        \vspace{1.ex}
        \label{lst:math}
    \end{subcaptionblock}
    \begin{subcaptionblock}{\textwidth}
        \begin{lstlisting}
            axm %mem.M:     *;
            axm %mem.Ptr:   * -> *;
            axm %mem.alloc: [T: *] [%mem.M] -> [%mem.M, %mem.Ptr T];
            axm %mem.free:  {T: *} [%mem.M, %mem.Ptr T] -> %mem.M;
            axm %mem.load:  {T: *} [%mem.M, %mem.Ptr T] -> [%mem.M, T];
            axm %mem.store: {T: *} [%mem.M, %mem.Ptr T, T] -> %mem.M;
            axm %mem.slot:  [T: *] [%mem.M, id: Nat] -> [%mem.M, %mem.Ptr T];
            axm %mem.lea:   {n: Nat, Ts: <<n; *>>} [%mem.Ptr <<j: n; Ts#j>>, i: Idx n] -> %mem.Ptr Ts#i;
        \end{lstlisting}
        \vspace{-1.ex}
        \caption{The \texttt{\%mem} plugin}
        \vspace{1.ex}
        \label{lst:mem}
    \end{subcaptionblock}
    \begin{subcaptionblock}{\textwidth}
        \begin{lstlisting}
            axm %autodiff.AD:          * -> *;              // marks type-level transformation
            axm %autodiff.ad: [T: *] [T] -> %autodiff.AD T; // marks term-level transformation
        \end{lstlisting}
        \vspace{-1.ex}
        \caption{The \texttt{\%autodiff} plugin}
        \vspace{-1.ex}
        \label{lst:autodiff}
    \end{subcaptionblock}
    \vspace{-4ex}
    \caption{Selection of plugins we implemented (excerpts). Normalizer information has been elided.}
    \label{sec:plugins}
\end{figure}
%    \begin{subcaptionblock}{\textwidth}
%        \begin{lstlisting}
%            axm %matrix.Mat:       [n m: Nat, T: *] -> *;
%            axm %matrix.unit:      [n m: Nat, T: *][%mem.M] -> [%mem.M, %matrix.Mat (n, m, T)];
%            axm %matrix.transpose: [n m: Nat, T: *][%mem.M, %matrix.Mat (n, m, T)]
%                                                    -> [%mem.M, %matrix.Mat (m, n, T)];
%            axm %matrix.sum:  [n m: Nat, [p e: Nat]]
%                [%mem.M, %matrix.Mat (n, m, %math.F (p, e))] -> [%mem.M, %math.F (p, e)];
%            axm %matrix.prod: [n m o: Nat, [p e: Nat]]
%                [%mem.M, %matrix.Mat (n, m, %math.F (p, e)), %matrix.Mat (m, o, %math.F (p, e))]
%              -> [%mem.M, %matrix.Mat (n, o, %math.F (p, e))];
%        \end{lstlisting}
%        \vspace{-1.ex}
%        \caption{The \texttt{\%matrix} plugin (simplified: the actual plugin supports n-ranked tensors)}
%        \vspace{-1.ex}
%        \label{lst:matrix}
%    \end{subcaptionblock}

\subsubsection{The core Plugin}
\label{sec:core_plugin}

This plugin (\autoref{lst:core}) defines arithmetic operations (\lstinline|%core.nat|) and comparisons (\lstinline|%core.ncmp|) for \lstinline|Nat| and integer operations on values of type \lstinline|Idx s|.
All integer operations abstract over the size~\lstinline|s|.
Like in \llvm, it is up to the operation to decide whether a specific \lstinline|Idx s|-typed value is signed or unsigned.
For example, the right-shift \lstinline|%core.shr| comes in two flavors: \lstinline|a|rithmetic (with sign extension), and \lstinline|l|ogical (without sign extension).
Integer operations like \lstinline|%core.wrap.add| take an overflow mode~\lstinline|m|. % and an \lstinline|Idx| size \lstinline|s| followed by the operands \lstinline|a| and \lstinline|b|.
This mode dictates how overflow is handled (wrap-around or undefined behavior for both signed and unsigned overflow).
Along the same lines, \lstinline|%core| introduces all 4 unary (\lstinline|%core.bit1|), all 16 binary bitwise (\lstinline|%core.bit2|), as well as the usual signed/unsigned comparison (\lstinline|%core.icmp|) operations.
The \lstinline|%core.conv| axiom converts between different-sized \lstinline|s|igned or \lstinline|u|nsigend values via truncation, zero-, or sign-extension whereas \lstinline|%core.bitcast| allows for arbitrary (potentially unsafe) casts.

For convenience, there is a function \lstinline|%core.minus| available that builds a unary minus by subtracting the given operand \lstinline|a| from \lstinline|0| (see also \autoref{sec:infer}-\ref{sec:cpp}).
This function will always be inlined due to the default \lstinline|tt| filter (\autoref{sec:filter}).
Note that in \llvm, \mlir, and similar \acp{IR} such helpers are usually implemented as C++ code that directly emit the desired code snippet as their type systems cannot express polymorphic/dependently typed functions.
The \mimir function \lstinline|%core.minus| on the other hand can be called and passed around just like any other axiom.

\subsubsection{The math Plugin}

This plugin (\autoref{lst:math}) introduces a type operator \lstinline|%math.F| that expects the number of significant precision bits~\lstinline|p| and exponent bits~\lstinline|e| over which all \lstinline|%math| operations abstract.
Additionally, most operations expect a mode~\lstinline|m| that fine-adjusts how strictly they should obey the IEEE-754 standard for floating-point transformations.
For example, you may choose to allow reassociation of floating-point operations or ignore \verb|NaN|s or infinity.
The axioms for the actual operations such as \lstinline|%math.arith|metic, \lstinline|%math.tri|gonometric, or floating-point comparisons (\lstinline|%math.cmp|) and convenience wrappers like \lstinline|%math.minus| are straightforward.

\begin{wraptable}{r}{.52\textwidth}
% \begin{table}
    \centering
    \vspace{-2.9ex}
    \begin{footnotesize}
    \begin{tabular}{lrrr}
        \toprule
        Benchmark       & \hspace{-8em}Input size    & C [s]              & \impala [s] \\
        \midrule
        aobench         & --            &  $0.667 \pm 0.014$ &  $0.659 \pm 0.004$ \\
        fannkuch        & 12            & $27.709 \pm 0.210$ & $26.301 \pm 0.186$ \\
        fasta           & \hspace{-8em}25,000,000    &  $0.711 \pm 0.011$ &  $0.814 \pm 0.013$ \\
        mandelbrot      & 5,000         &  $1.393 \pm 0.011$ &  $1.390 \pm 0.010$ \\
        meteor          & 2,098         &  $0.036 \pm 0.001$ &  $0.036 \pm 0.004$ \\
        nbody           & \hspace{-8em}50,000,000    &  $3.621 \pm 0.031$ &  $2.825 \pm 0.021$ \\
        pidigits        & 10,000        &  $0.371 \pm 0.005$ &  $0.370 \pm 0.004$ \\
        spectral        & 5,500         &  $1.387 \pm 0.011$ &  $1.409 \pm 0.014$ \\
        regex           & --            &  $4.101 \pm 0.037$ &  $4.128 \pm 0.044$ \\
        reverse         & --            &  $0.744 \pm 0.025$ &  $0.714 \pm 0.029$ \\
        \midrule
        geom.~speedup   & --            &  $1$               &  $1.020$           \\
        \bottomrule
    \end{tabular}
    \end{footnotesize}
    \caption{Original C version vs.~our \impala implementation that uses \mimir (lower is better)}
    \label{tab:performance}
    \vspace{-4ex}
% \end{table}
\end{wraptable}

\subsubsection{The mem Plugin}
\label{sec:mem_plugin}

This plugin (\autoref{lst:mem}) introduces a type \lstinline|%mem.M| to abstract from the machine state (\autoref{sec:overview_types}).
%\enquote{memory} type \lstinline|%mem.M| to abstract from the machine state
%Memory is handled linearly in the program.
%This means, there exists at most one memory value that is manipulated via the \lstinline|%mem| axioms.
%For this reason, all axioms that potentially may have side effects receive and produce a \lstinline|%mem.M| instance.
%This encoding is similar to the \lstinline|IO| monad in Haskell.% and Idris.\footnote{The underlying encoding in Idris is via a \lstinline|WorldVal| value analogously to our \lstinline|mem|.}.
We expose axioms to allocate memory, load from, and store values into previously allocated pointers, and to index into pointers that point to compound data types.
%Load, stores, and allocations are represented by the axioms of the same name.
Most axioms expect a machine state and additional arguments like the pointer to operate on, and return a memory instance together with the produced results like the loaded value.
The \lstinline|%mem.lea|%
\footnote{The name is inspired by the x86 assembly instruction and its semantics is similar to \texttt{getelementptr} in \llvm but arguably more streamlined.}
axiom performs pointer arithmetic.
Since this instruction does not have side effects as it does not directly interact with memory, no machine state is needed.
The \lstinline|%mem.lea| axiom takes a pointer to a tuple or array and the offset \lstinline|i| in which it wants to index.
The result is the pointer to the \lstinline|i|-th element.
% mem: Mem, p: ptr Foo = alloc Foo mem;
\begin{example}\label{ex:lea}
    In the following listing, \lstinline|p1| points to a 3-tuple and \lstinline|%mem.lea| indexes into the second element ($1_3$).
    Then, \lstinline|%mem.lea| computes the pointer to the \lstinline|i|$^\mathit{th}$ element of an array of size \lstinline|n|.
    Both \lstinline|i| and \lstinline|n| are statically unknown.
    In particular, note how \lstinline|%mem.lea|'s dependent type computes with the help of the normalization rules the result type of the element pointers.
    \begin{lstlisting}
        let p1/*: %mem.Ptr [Nat, I16, Nat]*/ = /*...*/;
        let q1/*: %mem.Ptr I16              */ = %mem.lea /*(3, (Nat, I16, Nat)*/ (p1, 1_3);
        let i /*: Idx n                     */ = /*...*/;
        let p2/*: %mem.Ptr <<n; Nat>>       */   = /*...*/;
        let q2/*: %mem.Ptr Nat              */ = %mem.lea /*(3, <n; Nat>)*/ (p2, i);
    \end{lstlisting}
\end{example}

%Additionally, this plugin also contains the code for the SSA and copy propagation pass as outlined in \autoref{sec:rewriting:opt}
%In the plugin, we introduce a backend pass that translates the program code into \llvm \ac{IR}.

\subsubsection{Evaluation}
\label{sec:impala}

In this experiment, we want to confirm that the low-level plugins perform as well as directly using \llvm.
To this end, we ported the code generator of the research language \impala~\cite{DBLP:conf/cgo/LeissaKH15} to \mimir~2.
We ported the fastest available C implementations that were neither manually vectorized nor parallelized from \emph{The Computer Language Benchmarks Game}~\cite{benchmarksgame} to \impala.
In addition, we ported the publicly available \texttt{aobench}~\cite{aobench}.
\autoref{tab:performance} shows that the performance is as expected nearly identical to the original C versions (with two slight outliers---one for the better, the other one for the worse).
Note that the \texttt{regex} benchmark does \emph{not} use the \texttt{\%regex} plugin.

\subsection{Regular Expressions}
\label{sec:eval:regex}

We have already discussed the \texttt{\%regex} plugin in \autoref{sec:overview}:
%As a reminder, this is a simple use case that showcases the power of the plugin framework present in \mimir.
The plugin defines a set of axioms representing ranges of literals, consecutive elements (conjunction), alternative elements (disjunction), negation, and quantifiers.
These are sufficient to define a useful set of compile-time regular expressions.
While we only test this plugin with \mim, one could easily integrate \mimir with this plugin as a code generator into a \ac{RegEx} parser and either generate code for a \ac{RegEx} at compile time or \ac{JIT}-compile at run time.
%This can be achieved either in an offline or \ac{JIT} compiled fashion.
%We provide the typical classes, such as \verb|\w| for word characters and \verb|\d| for digits, as mere definitions of alternative ranges.
%For example, \verb|\w| is implemented as \verb|[a-zA-Z0-9_]|.
%Literals are defined as the range \verb|[a-a]|.

% if we need space: remove this, the example was in overview already:
%As an example, consider the following \ac{RegEx} that matches simple top-level domains: \verb|^\w+\.\w+$|.
%Using the axioms in \mimir, this is defined as:
%\begin{lstlisting}
%%regex.conj (%regex.quant.plus %regex.cls.w, %regex.lit '.', %regex.quant.plus %regex.cls.w)
%\end{lstlisting}

% When comparing to other \ac{RegEx} engines, it has to be taken into account that our implementation does lack a number of features that the other engines do provide:
% Currently not supported in \mimir's plugin is the parsing of \ac{RegEx} pattern strings as well as capturing groups.

This showcase demonstrates that \mimir provides a powerful core whose extensibility indeed makes implementing a DSL with intrinsic, normalizable expressions and domain-specific optimizations very straightforward.
Our \texttt{\%regex} plugin provides a legalization pass that is written in C++ and receives the normalized \ac{RegEx} pattern in its opaque form.
This allows the pass to use the exhaustive understanding of the \ac{RegEx} to generate an optimized matcher using finite automatons.
To do so, the pass first translates the \ac{RegEx} into \iac{NFA}, further converts this to a \ac{DFA} \cite[2.2]{DBLP:books/daglib/0031526}, and then minimizes it~\cite{hopcroft1971n}.
Finally, the minimal \ac{DFA} is translated into low-level control flow in \mimir's IR that is purely based on integer comparisons as well as jumps between state continuations.
The generated code is put into action by the pass as it replaces the pattern application with a call to the newly generated matcher function.

\begin{wraptable}{r}{.45\textwidth}
    \centering
    \vspace{-2.9ex}
    \begin{footnotesize}%
    \begin{tabular}{lrrr}
        \toprule
        RegEx       &                           & \multicolumn{1}{c}{Compile} & \multicolumn{1}{c}{Match}  \\
        Engine      & LoC                       & time [ms]             & time [µs] \\
        \midrule
        CTRE        & 4,153                     & 1,677                 & 4,736 \\
        std::regex  & 4,874                     & 2,207                 & 10,151 \\
        pcre2       & \multirow{2}{*}{85,879}   & \multirow{2}{*}{67}   & 3,882 \\
        pcre2-jit   &                           &                       & 1,308 \\
        hand-written C\hspace{-1em}   & 102                       & 45                    & 886 \\
        \mimir     & 941                       & 145                   & 640 \\
        \bottomrule
    \end{tabular}
    \end{footnotesize}%
    \caption{Comparison of \ac{RegEx} engines}
    \vspace{-3ex}
    \label{tab:regex-engines}
\end{wraptable}

In total, the \ac{RegEx} plugin only encompasses 919 lines of C++ code and 22 lines of \mimir code.
We compare our \ac{RegEx} implementation in \mimir with \ac{CTRE}, \verb|std::regex|~\cite{stdregex}, as well as \ac{PCRE2} in an interpreted and a \ac{JIT} compiled variant.
From the \ac{LoC} numbers in \autoref{tab:regex-engines} we observe that \mimir's implementation is at least one order of magnitude less complex.
Despite the rather low complexity, our engine outperforms state-of-the-art \ac{RegEx} engines.
It is even 28\% faster than a manually written, low-level matcher that we implemented in C and matches only the specific pattern below.
Most likely, our C implementation contains some redundant checks, but spotting them is very hard in such low-level code that comes from manually writing a complicated matcher.

The listed \ac{LoC} include actual code lines exclusively.\footnote{\label{fn:cloc}\ac{LoC} as reported by \href{https://github.com/AlDanial/cloc}{cloc}, treating \mimir code as Rust}
\enquote{Compile time} is the execution time of \verb|clang++| and for \mimir the \mimir frontend and optimizer.
All benchmarks test the following \ac{RegEx} on 10,215 E-Mail addresses that are matched by this pattern and 450 that are not~\cite{clairFraudEMails}:
\begin{lstlisting}[breaklines=true,basicstyle=\ttfamily\tiny,mathescape]
    ^[a-zA-Z0-9](?:[a-zA-Z0-9]*[._\-]+[a-zA-Z0-9])*[a-zA-Z0-9]*@[a-zA-Z0-9](?:[a-zA-Z0-9]*[_\-]+[a-zA-Z0-9])*[a-zA-Z0-9]*\.(?:(?:[a-zA-Z0-9]*[_\-]+[a-zA-Z0-9])*[a-zA-Z0-9]+\.)*[a-zA-Z][a-zA-Z]+$\$$
\end{lstlisting}
We have also tested other regular expressions with similar results.

% vim:spell:spelllang=en
\subsection{Machine Learning}\label{sec:ml}

\subsubsection{Tensor Plugin}\label{sec:tensor}

%This basically boils down
%In all of our case studies, however, we did not encounter an issue in the sense that we wanted to write down a program that \mimir's type checker refused to accept.
%Normalization coupled with partial evaluation worked reliably in our experience.
%Keep in mind that \mimir does not want to be a proof assistant.
%If everything else fails, a programmer can simply insert a (potentially unsafe) cast via an unsafe axiom like \lstinline|%core.bitcast| (see also \autoref{sec:meta}).

The \verb|%tensor| plugin provides operations that abstract over the tensors' rank, shape, and element type.
The rank and shape of involved tensors is tracked in their types across operations.
For instance, the matrix product takes two matrices of size $l \times m$ and $m\times n$ returning a matrix of size $l \times n$.
Note that $l$, $n$, and $m$ do not necessarily have to be constants---a common restriction in many other programming idioms/systems including XLA~\cite{xla} or C++ templates.
% Dimensions with a size of \hl{1} are normalized away leaving a scalar element.
Operations like getting the shape of a tensor are statically optimized away by directly accessing the information of the type.
Additionally, \mimir's type system guarantees that read and write accesses occur within bounds, eliminating the need for dynamic bound checks.

Using \mimir's expressive type system, the \verb|%tensor| plugin defines a general \hl{map\_reduce} function in the spirit of Numpy's \href{https://numpy.org/doc/stable/reference/generated/numpy.einsum.html}{\texttt{einsum}} function.
The plugin eventually lowers all other tensor operations into a \hl{map\_reduce} call.
This significantly cuts the number of cases to check in optimizations.
In particular, many peephole optimizations like collapsing two consecutive transpositions into the identity are handled automatically.
Furthermore, the usage of high-level axioms also benefits other plugins.
For example, the \verb|%autodiff| plugin (see below) directly operates on the high-level axioms of the \verb|%tensor| plugin.

As a non-trivial example that showcases many of \mimir's features, consider the following axiom in the spirit of XLA's \href{https://openxla.org/xla/operation_semantics\#variadic_reduce}{variadic reduction}:
\begin{lstlisting}
axm %tensor.reduce: {r: Nat, s: <<r; Nat>>, ni: Nat, Is: <<ni; *>>}
                    [f: <<2; <<i: ni; Is#i>> >> -> <<i: ni; Is#i>>]
                    [is: <<i: ni; <<s; Is#i>> >>, init: <<i: ni; Is#i>>, dims: <<r; Bool>>]
                  -> <<i: ni; << <j: r; (s#j, 1)#(dims#j)>; Is#i>> >>;
\end{lstlisting}
This operation is polymorphic in the number of inputs \hl{ni}, the rank \hl{r} and the shape \hl{s} of the involved tensors, as well as their element types~\hl{Is}.
The reduction function \hl{f} expects a pair of \hl{ni}-tuples whose elements correspond to \hl{Is} and yields again such an \hl{ni}-tuple.
The \hl{ni}-tuple \hl{init} comprises the initial values of the reduction.
The Boolean mask vector \hl{dims} of size \hl{r} selects for each dimension whether it should be reduced.
For this reason, the reduction yields \hl{ni} arrays of rank \hl{r} whose shape is computed by either using the original dimension or \hl{1}, if the corresponding element in the \hl{dims} vector is \lstinline|tt|:
Since dimensions of arity~\hl{1} will collapse during normalization, the resulting array will only exhibit dimensions that were not selected with \hl{dims}.
For example, suppose \hl{a}, \hl{b}, and \hl{c} are $n \cdot 4 \times n$ matrices with element types \lstinline|I8|, \lstinline|I16|, and \lstinline|I32|, respectively.
% Furthermore, suppose function \hl{f} has type \hl{[\key{I8}, \key{I16}, \key{I32}] $\to$ \key{I32}}.
Then, the following expression
\begin{lstlisting}
lam f (x y: [I8, I16, I32]): [I8, I16, I32] = /*...*/;
let n4  = %core.nat.mul (n, 4);
%tensor.reduce /*(2, (n4,n), 3, (I8,I16,I32))*/ f ((a, b, c) (0I32, 0I16, 0I32) (ff, tt))
\end{lstlisting}
has type \hl{[<<n4; \key{I8}>>, <<n4; \key{I16}>> ,<<n4; \key{I32}>>]}.
Note that although the matrix dimensions are \emph{not} compile-time constants, the dependent types keep track of the information that the rows of the input matrices as well as the lengths of the output vectors are of size \hl{n4}.
This information may be useful during vectorization, for instance, to remove edge cases.
You can match applications of this axiom from within C++ for further analyses/optimizations:
\begin{lstlisting}[language=C++,literate=]
if (auto reduce = match<tensor::reduce>(e)) { auto [is, init, dims] = reduce->args(); /*$\color{codegreen}\mydots$*/ }
\end{lstlisting}
    % auto f                = reduce->decurry()->arg();
    % auto [r, s, ni, Is]   = reduce->decurry()->decurry()->args();
    % // ...
% Note once again that \lstinline|let|-expressions do not exist in \mimir's graph.
% Thus, scrutinizing the arity of the output vectors will directly yield \hl{\%core.nat.mul (n, 4)}.

\subsubsection{Autodiff Plugin}\label{sec:autodiff}

\renewcommand{\thornado}{\texttt{\%autodiff}}

% \todo[inline]{encorporate with other sections (conclusion and where eval is mentioned)}
% \todo[inline]{shorten section a bit}
% \todo[inline]{show AD type}
%\todo[inline]{\thornado -> autodiff}
%\todo[inline]{1 sentence what autodiff}
The \lstinline|%autodiff| plugin implements reverse-mode \acf{AD}---a prominent technique to compute the derivatives of code.
These derivatives are used to optimize parameters using gradient descent methods in, for example, machine learning frameworks.

% \todo[inline]{1 paragraph what we did}

% high-level transformation -> rewrite
% continuation nesting
% use example from intro

\lstset{escapeinside={<@}{@>}}
\begin{figure}
    \centering
    \begin{minipage}{.3\linewidth}
    \begin{lstlisting}
    fun (x: T, y: T): T =
      let z = x + y;
      return (x * z);
    \end{lstlisting}%
    % \begin{lstlisting}
    % $\lambda$ (x,y).
    %   let z = x+y in
    %   x * z
    % \end{lstlisting}%
    \end{minipage}%
    \begin{minipage}{.7\linewidth}
    \begin{lstlisting}
    fun ((x, <@\texttt{\textcolor{ACMBlue}{x$^*$}}@>): [T, T -> T], (y, <@\texttt{\textcolor{ACMBlue}{y$^*$}}@>): [T, T -> T]): [T, T -> T] =
      let z, <@\texttt{\textcolor{ACMBlue}{z$^*$}}@> = (x + y, <@\texttt{\textcolor{ACMBlue}{fn (s: T): T = return (x$^*$(s) + y$^*$(s))}}@>);
      return (x * z, <@\texttt{\textcolor{ACMBlue}{fn (s: T): T = return (x$^*$(z*s) + z$^*$(x*s))}}@>);
    \end{lstlisting}
    % \begin{lstlisting}
    % $\lambda$ ((x, <@\texttt{\textcolor{ACMBlue}{x$^*$}}@>), (y, <@\texttt{\textcolor{ACMBlue}{y$^*$}}@>)).
    %   let z, <@\texttt{\textcolor{ACMBlue}{z$^*$}}@> = (x+y, <@\texttt{\textcolor{ACMBlue}{$\lambda$s. x$^*$(s) + y$^*$(s)}}@>) in
    %   (x * z, <@\texttt{\textcolor{ACMBlue}{$\lambda$s. x$^*$(z*s) + z$^*$(x*s)}}@>)
    % \end{lstlisting}
    \end{minipage}
    \vspace{-2ex}
    \caption{Each computation of the original program (left) is augmented with a backpropagator marked with $\dashedph[x]^*$ (colored in blue, right). The differentiated function returns $x\cdot (x+y), \lambda s. s\cdot (2x+y, x)$. \mimir-like pseudocode.}
    \label{fig:pb_func}
    \vspace{-2ex}
\end{figure}

A popular way to implement reverse-mode \ac{AD} in functional languages is to use \emph{backpropagators}~\cite{pearlmutter2008reverse}.
Each function~$f$ is augmented to return a function $f^\ast$, the backpropagator, in addition to its original result (see \autoref{fig:pb_func} for an example):
$D\,f\,(x,x^\ast)\coloneqq \left(f\,(x),\lambda a.x^\ast\,(f'\,(x)\cdot a)\right)$.

By this technique, \ac{AD} effectively builds a list of backpropagators:
The backpropagator of~$f$ uses~$x^\ast$ which is the backpropagator of the function that was used to compute the argument of~$f$.
In turn, $f$'s backpropagator is passed to all functions that take the value $f$~returned as an argument.
The derivative is then computed by invoking the backpropagator of the end result with~$1$.
%The backpropagator is a function that takes the derivative with respect to the function result as well as the adjoint and returns the derivatives with respect to the function inputs. % , the tangent. \todo[inline]{\enquote{, the tangent?} what?}
%The formulation of gradients as pullback allows encoding the chain rule as function composition.
% The backpropagator returns the derivatives in the direction passed as its argument. By doing so, the composition of backpropagators constitutes the application of the chain rule.
% CPS -> reversal => too complicated for one sentence
% ^* erklären
% The backpropagator computes the derivatives of the corresponding function scaled with the argument of the backpropagator.
%Doing so, backpropagators incorporate the chain rule and

The \texttt{\%autodiff} plugin (\autoref{lst:autodiff}) expresses this transformation as a set of local rewrite rules that are applied in a bottom-up fashion to transform the original function into its derivative.
Besides simple functions, the \texttt{\%autodiff} plugin handles a rich set of features including non-scalar types, pointers, and higher-order, recursive functions.
%The advantage of the functional higher-order approach to \ac{AD} using pullbacks is that it is simple to implement and modular in the sense that \ac{AD} of a compound language element translates to differentiating its constituents.
The advantage of the local-rewrite approach to \ac{AD} is that it is simple to implement and modular in the sense that \ac{AD} of a compound language element translates to differentiating its constituents.
This modularity allows us to extend the implementation at a high level by just specifying the derivative of another operation.
For example, the \texttt{\%tensor} plugin specifies derivatives for common matrix operations without tainting the \texttt{\%autodiff} plugin with dependencies on the \texttt{\%tensor} plugin.
By doing so, the plugin computes the derivative of a matrix product using matrix products instead of low-level for-loops.
The use of optimal code generation for these high-level operations results in a significant speedup of the resulting program---as demonstrated by \citet{peng2023lagrad}.

The use of higher-order functions in the derivative computation makes it more challenging for the compiler to produce efficient code.
However, \mimir's optimizer and partial evaluator are able to remove this overhead entirely in the benchmarks that we considered as our evaluation below shows.
In particular, \mimir's design resolves many practical implementation problems basically for free:
Most notably, hash-consing merges identical backpropagator invocations, and partially evaluating backpropagators removes most of the boilerplate that is introduced by the newly introduced nested function calls.
% partial eval opt -> less function calls, no nesting, resolve boilerplate resulting from modularity

% The sentence is not meaningful without the code below -> also remove it
% The \texttt{\%autodiff} plugin (\autoref{lst:autodiff}) exposes an axiom to transform functions into the differentiated version (\autoref{}). % as well as interfaces for the compilation passes to perform the actual differentiation.
% \todo[inline]{The last sentence is unclear. What specifically?}
%\begin{lstlisting}
%axm %autodiff.AD:            * -> *             , normalize_AD; // type-level transformation
%axm %autodiff.ad: PI [T: *] -> T -> %autodiff.AD T, normalize_ad;

%axm %autodiff.ad_eval_pass: %compile.Pass;
%\end{lstlisting}

To show that \mimir is able to succinctly optimize the code that results from applying backpropagator-based \ac{AD}, we compare it against the state-of-the-art \ac{AD} frameworks \pytorch~\cite{pytorch} and \enzyme~\cite{moses2020instead}.
\pytorch builds dynamic computation graphs via operator overloading in Python.
\enzyme differentiates \llvm code during compilation.

% We find that \mimir's optimizations effectively remove the boilerplate code and optimize the resulting gradient computation.
%We compare our implementation to the popular frameworks \pytorch{} \cite{pytorch_NEURIPS2019_9015} and \enzyme~\cite{moses2020instead}.

%Although our approach is more similar to other high-level approaches known from functional programming languages from a theoretical point of view, we compare \thornado{} to the low-level implementations of state-of-the-art \ac{AD}. % the faster performance of
%We show that our modular functional approach competes with modern highly optimized implementations due to compiler optimizations that remove the boilerplate code and optimize the resulting gradients.

\paragraph{Setup}

% In our benchmark suite we use the Microsoft \adbench suite \cite{srajer2018benchmark}, simple matrix operations, and a feedforward neural network for the MNIST dataset ().

We evaluate the frameworks on Microsoft's \adbench~suite~\cite{srajer2018benchmark}.
We use the Gaussian mixture model (GMM), bounded analysis (BA), and a long short-term memory neural network (LSTM) from the \adbench~suite.
% To compare with \enzyme, we also use their test case for the Fast Fourier Transformation (FFT) and for the simulation of the Brusselator system~\cite{feinberg1987chemical,yu2018mathematical}. %  on matrix operations
Additionally, we compare the approaches on a network for the MNIST classification task~\cite{deng2012mnist}.
For a fair comparison to \enzyme, we instruct the \texttt{\%tensor} plugin to generate low-level, straightforward loop nests.
Alternatively, the plugin could also employ highly tuned \blas~\cite{blackford2002updated} routines, instead.
But we are interested in how well the \texttt{\%autodiff} plugin copes with low-level loop nests as this abstraction layer corresponds to the knowledge \enzyme obtains via \llvm's scalar evolution analyses.

\subsubsection{Running Time}
\label{sec:eval_rt}

% Timing
%\begin{wrapfigure}{r}{0.7\textwidth}
\begin{figure}
    \centering
    % \vspace{-3ex}
    % .49
        % \input{fig/benchmark_relativ_shared}
    % \begin{subcaptionblock}{.6\textwidth}
    %     \input{fig/benchmark_relativ_enzyme_only}
    %     % \vspace{-1.ex}
    %     % \caption{ABC}
    % \end{subcaptionblock}
    % \begin{subcaptionblock}{.6\textwidth}
    %     \input{fig/benchmark_relativ_pytorch_only}
    %     % \vspace{-1.ex}
    %     % \caption{DEF}
    % \end{subcaptionblock}
    % \input{fig/benchmark_relativ}
    \begin{tikzpicture}% [trim axis left]
\begin{axis}[
	  x tick label style={
            % transparent, font=\tiny
	 	font=\footnotesize,
        },
    xtick = data,
    % xtick=\empty, %axis line style=transparent,
	ylabel=Speedup,
	ybar,
	bar width=7pt,
    width=.40\linewidth,
    height=4cm,
    symbolic x coords={
        0,
        {GMM$_1$  },
        {GMM$_2$  },
        {BA$_1$   },
        {BA$_2$   },
        {MNIST$_1$},
        {MNIST$_2$},
        {LSTM     },
        1
    },
    xticklabel style={rotate=90},
    nodes near coords,
    nodes near coords align={horizontal},
    % enlarge y limits=1,
    % nodes near coords align={top},
    every node near coord/.append style={font=\scriptsize,anchor=south},%, yshift=10mm},
    point meta=rawy,
    ymax = 60000,
    ymode=log,
    legend style={at={(0.52,0.98)},anchor=north west,font=\footnotesize},
    label style={font=\small},
    %
    % y tick label style={
    %     /pgf/number format/.cd,
    %         fixed,
    %         fixed zerofill,
    %         precision=0,
    %     /tikz/.cd
    % },
    scale only axis
]

% \addplot [fill=chartgreen] % Enzyme
% 	coordinates {
% ({GMM$_1$  }, 1.33)
% ({GMM$_2$  }, 1.09)
% ({BA$_1$   }, 0.68)
% ({BA$_2$   }, 0.64)
% ({MNIST$_1$}, 0.86)
% ({MNIST$_2$}, 0.95)
% ({LSTM       }, 1.13)
%     };
%    \addlegendentry{\enzyme};

\addplot [fill=chartyellow%, point meta=explicit symbolic,
            % nodes near coords,
            % visualization depends on={y \as \myy},
            % nodes near coords style={at={(0,0)}},
            % every node near coord/.append style={font=\scriptsize,at={(0,0)}},
            % visualization depends on=y \as \rawy,
            % every node near coord/.append style={
            %         font=\tiny,
            %         shift={(axis direction cs:0,-2)}
            %     }
    %point meta=rawy,
    %nodes near coords={
    %    %\pgfmathtruncatemacro\nValue{\coordindex+1}%
    %    %$\nValue \theta$%
    %    \contour{white}{\pgfplotspointmeta}
    %},
] % PyTorch
	coordinates {
({GMM$_1$  }, 5.09) [5.09]
({GMM$_2$  }, 3.13) [3.13]
({BA$_1$   }, 3171.29) [\contour{white}{3171.29}]% \contour{white}{3171.29}
% ({BA$_2$   }, nan)% nan
({MNIST$_1$}, 1.24) [1.24]
({MNIST$_2$}, 3.62) [3.62]
({LSTM       }, 146.73) [146.73]
    };
   \addlegendentry{\pytorch};

\addplot[gray,dashed,sharp plot,update limits=false]
	coordinates {
        (0, 1.0)
        (1, 1.0)
    };

\end{axis}
\end{tikzpicture}
\vspace{-1ex}
    \hfill
    \begin{tikzpicture}[trim axis left]
\begin{axis}[
	  x tick label style={
	 	font=\footnotesize,
        },
    xtick = data,
	% ylabel=Speedup,
    % ymin = 0,
	ybar,
	bar width=7pt,
    width=.40\linewidth,
    height=4cm,
    symbolic x coords={
        0,
        {GMM$_1$  },
        {GMM$_2$  },
        {BA$_1$   },
        {BA$_2$   },
        {MNIST$_1$},
        {MNIST$_2$},
        {LSTM       },
        1
    },
    xticklabel style={rotate=90},
    nodes near coords,
    nodes near coords align={horizontal},
    % enlarge y limits=1,
    % nodes near coords align={top},
    every node near coord/.append style={font=\scriptsize,below},%, yshift=10mm},
    point meta=rawy,
    ymax = 2,
    ymode=log,
    legend style={at={(0.32,0.98)},anchor=north west,font=\footnotesize},
    label style={font=\small},
    %
    % y tick label style={
    %     /pgf/number format/.cd,
    %         fixed,
    %         fixed zerofill,
    %         precision=0,
    %     /tikz/.cd
    % },
    scale only axis
]

\addplot [
    fill=chartgreen,
    coordinate style/.condition={\coordindex==0}{yshift=4mm,anchor=center},
    coordinate style/.condition={\coordindex==1}{yshift=4mm,anchor=center},
    coordinate style/.condition={\coordindex==6}{yshift=4mm,anchor=center},
    ] % Enzyme
	coordinates {
({GMM$_1$  }, 1.33)
({GMM$_2$  }, 1.09)
({BA$_1$   }, 0.68)
({BA$_2$   }, 0.64)
({MNIST$_1$}, 0.86)
({MNIST$_2$}, 0.95)
({LSTM       }, 1.13)
    };
   \addlegendentry{\enzyme};

% \addplot [fill=chartyellow, point meta=explicit symbolic,
%             % nodes near coords,
%             % visualization depends on={y \as \myy},
%             % nodes near coords style={at={(0,0)}},
%             every node near coord/.append style={font=\scriptsize,at={(0,0)}},
%             % visualization depends on=y \as \rawy,
%             % every node near coord/.append style={
%             %         font=\tiny,
%             %         shift={(axis direction cs:0,-2)}
%             %     }
%     %point meta=rawy,
%     %nodes near coords={
%     %    %\pgfmathtruncatemacro\nValue{\coordindex+1}%
%     %    %$\nValue \theta$%
%     %    \contour{white}{\pgfplotspointmeta}
%     %},
% ] % PyTorch
% 	coordinates {
% ({GMM$_1$  }, 5.09) [5.09]
% ({GMM$_2$  }, 3.13) [3.13]
% ({BA$_1$   }, 3171.29) [\contour{white}{3171.29}]% \contour{white}{3171.29}
% ({BA$_2$   }, nan)% nan
% ({MNIST$_1$}, 1.24) [1.24]
% ({MNIST$_2$}, 3.62) [3.62]
% ({LSTM       }, 146.73) [146.73]
%     };
%    \addlegendentry{\pytorch};

\addplot[gray,dashed,sharp plot,update limits=false]
	coordinates {
        (0, 1.0)
        (1, 1.0)
    };

\end{axis}
\end{tikzpicture}
\vspace{-1ex}
    \vspace{-1ex}
    \caption{Speedup $\nicefrac{t}{t_{\text{mim}}}$ of \mimir vs.~\enzyme, and \pytorch\vspace{-1ex}}
    % \caption{Speedup $\nicefrac{t}{t_{\text{mim}}}$ of \mimir vs.~\enzyme. \pytorch is not shown due to the much slower runtime}
    \label{fig:benchmark}
    \vspace{-2ex}
\end{figure}
% \end{wrapfigure}
We ran the benchmarks using a single thread on an AMD Ryzen 7 5800X with 32 GB of DDR4@2400MT/s RAM.
%We used \llvm/Clang 15.0.7 for the C sources as well as the emitted \llvm code.
\autoref{fig:benchmark} compares the speedup/slowdown of \enzyme/\pytorch compared to \texttt{\%autodiff} as baseline.
For the \mimir{} implementations, we write the benchmarks in Impala (\autoref{sec:impala}) that compiles to \mimir{} and uses the \thornado{} \ac{AD} compiler pass.
Finally, \mimir{} emits an \llvm~file that we feed to \clang~to generate an executable.
\enzyme is an \llvm pass written in C++ while \pytorch is written in Python.

Our evaluation shows that \enzyme{} and \thornado{} are comparable in performance.
In some cases like bounded analysis, \enzyme has a better caching/recomputation balance resulting in lower runtime.
In other cases like the GMM or LSTM benchmark, \thornado{} is faster due to more caching.
The main performance difference is due to caching or recomputation of intermediate results.
\enzyme{} manages to better detect induction variables and recomputes them in the backward pass instead of storing them.
On the other hand, it sometimes stores additional unnecessary intermediate results in the forward pass.

\begin{wraptable}{r}{.5\textwidth}
    \vspace{-2.9ex}
    \begin{center}
    \begin{footnotesize}
    \begin{tabular}{llrrr}
        \toprule
                                    &         &         \multicolumn{3}{r}{Cyclomatic complexity}             \\
                                    &         & LoC [k] & total & avg. \\
        \midrule
        \multirow{2}{*}{\enzyme}    & w/      & 58.9    & \phantom{XXX}13,793 & 10.9 \\
                                    & w/o     & 49.4    & 11,857 & 12.8 \\
        \multirow{2}{*}{\thornado}  & w/      &  4.7    &    704 &  \phantom{XXX}1.6 \\
                                    & w/o     &  1.3    &    141 &  1.8 \\
        \bottomrule
    \end{tabular}
    \end{footnotesize}
    \end{center}
    \begin{scriptsize}
        w/: with extensions \& plugins \hfill w/o: without extensions \& plugins
    \end{scriptsize}
    \caption{\Ac{LoC} and cycl. complexity measure (lower is better). We compute the measure per function; \emph{total} cyclomatic complexity refers to the sum over all files.}
    \label{tab:cyclomatic}
    \label{tab:loc_cyclomatic}
    \vspace{-3ex}
\end{wraptable}

In our evaluation, \pytorch{} is slower than \enzyme{} and \thornado.
This is partly due to the overhead of constructing the backward graph at runtime,
the overhead of the Python interpreter, more memory usage, and missing optimizations like inlining.
% \enzyme = mimir (sometimes slower => more caching)
% pytorch in our tests slower (some due to overhead)
%\pytorch seems to be especially inefficient in the bounded analysis benchmark.
\pytorch is inefficient in the bounded analysis benchmark resulting in a significantly longer running time.
This slowdown is caused by the deeply nested function calls and loops that contain conditionals in the benchmark source code.
 The $BA_2$ test timed out, and thus, is not listed in \autoref{fig:benchmark}.
%\todo[inline]{which deeply nested function calls?}

\begin{figure}
    \centering
    \begin{tikzpicture}
\begin{loglogaxis}[
	% legend style={at={(0.5,-0.15)},
	% 	anchor=north,legend columns=-1},
    %legend pos=south east,
    %legend style={
        %font=\footnotesize,
        %legend columns=3,
        %anchor=north,
        %at={(0.5,-0.1)}
    %},
    legend style={at={(0.02,0.98)},anchor=north west,font=\footnotesize,legend columns=3},
    xticklabel pos=upper,
    xlabel=Size,
    ylabel={Time [ms]},
    x label style={anchor=north west,xshift=-6.3cm},
    mark size=1pt,
    width=\linewidth,
    height=0.45\linewidth,
    grid=major,
    grid style={black!20, dashed},
    every axis plot/.append style={thick,mark size=2pt},
]

\addplot[color=chartred,mark=*] coordinates {
  (30, 160)
  (60, 170)
  (150, 190)
  (300, 216)
  (330, 182)
  (600, 254)
  (660, 199)
  (1155, 196)
  (1200, 357)
  (1650, 242)
  (2310, 221)
  (2805, 204)
  (3300, 332)
  (5610, 251)
  (5775, 331)
  (6600, 525)
  (10725, 253)
  (11550, 470)
  (13200, 893)
  (14025, 410)
  (21450, 377)
  (23100, 763)
  (28050, 666)
  (41925, 395)
  (46200, 1438)
  (53625, 679)
  (56100, 1144)
  (83850, 590)
  (107250, 1180)
  (112200, 2194)
  (209625, 1230)
  (214500, 2201)
  (419250, 2183)
  (429000, 4170)
  (838500, 4242)
};
   \addlegendentry{\torchscript};

\addplot[color=chartyellow,mark=x] coordinates {
  (30, 117)
  (60, 126)
  (150, 141)
  (300, 161)
  (330, 137)
  (600, 203)
  (660, 166)
  (1155, 156)
  (1200, 304)
  (1650, 213)
  (2310, 182)
  (2805, 164)
  (3300, 423)
  (5610, 217)
  (5775, 394)
  (6600, 724)
  (10725, 245)
  (11550, 679)
  (13200, 1329)
  (14025, 549)
  (21450, 485)
  (23100, 1187)
  (28050, 1013)
  (41925, 499)
  (46200, 2442)
  (53625, 1040)
  (56100, 1993)
  (83850, 935)
  (107250, 2059)
  (112200, 3920)
  (209625, 2253)
  (214500, 4043)
  (419250, 4394)
  (429000, 8434)
  (838500, 9163)
};
   \addlegendentry{\pytorch};

\addplot[color=chartorange,mark=o] coordinates {
  (30, 140)
  (60, 142)
  (150, 157)
  (300, 171)
  (330, 159)
  (600, 198)
  (660, 167)
  (1155, 161)
  (1200, 260)
  (1650, 200)
  (2310, 187)
  (2805, 178)
  (3300, 269)
  (5610, 219)
  (5775, 267)
  (6600, 382)
  (10725, 218)
  (11550, 373)
  (13200, 615)
  (14025, 343)
  (21450, 306)
  (23100, 598)
  (28050, 511)
  (41925, 319)
  (46200, 1064)
  (53625, 531)
  (56100, 873)
  (83850, 470)
  (107250, 928)
  (112200, 1650)
  (209625, 926)
  (214500, 1722)
  (419250, 1760)
  (429000, 3360)
  (838500, 3363)
};
   \addlegendentry{\pytorch 2.0 Multi-Core};

\addplot[color=chartpuruple,mark=+] coordinates {
  (30, 134)
  (60, 137)
  (150, 156)
  (300, 179)
  (330, 150)
  (600, 233)
  (660, 170)
  (1155, 166)
  (1200, 323)
  (1650, 224)
  (2310, 196)
  (2805, 182)
  (3300, 378)
  (5610, 238)
  (5775, 344)
  (6600, 639)
  (10725, 248)
  (11550, 560)
  (13200, 1233)
  (14025, 478)
  (21450, 381)
  (23100, 1071)
  (28050, 766)
  (41925, 410)
  (46200, 2136)
  (53625, 833)
  (56100, 1751)
  (83850, 761)
  (107250, 1858)
  (112200, 3662)
  (209625, 2065)
  (214500, 3832)
  (419250, 4298)
  (429000, 13818)
  (838500, 13625)
};
   %\addlegendentry{\pytorch 2.0 Single Core};
   \addlegendentry{\pytorch 2.0};

\addplot[color=chartgreen, mark=square*] coordinates {
  (30, 2)
  (60, 4)
  (150, 12)
  (300, 28)
  (330, 7)
  (600, 53)
  (660, 16)
  (1155, 16)
  (1200, 103)
  (1650, 41)
  (2310, 35)
  (2805, 31)
  (3300, 86)
  (5610, 64)
  (5775, 79)
  (6600, 171)
  (10725, 83)
  (11550, 181)
  (13200, 346)
  (14025, 151)
  (21450, 159)
  (23100, 351)
  (28050, 330)
  (41925, 255)
  (46200, 761)
  (53625, 444)
  (56100, 686)
  (83850, 526)
  (107250, 892)
  (112200, 1372)
  (209625, 1370)
  (214500, 1814)
  (419250, 2723)
  (429000, 3615)
  (838500, 5509)
};
   \addlegendentry{\enzyme};

\addplot[color=chartldarkblue, mark=triangle*] coordinates {
  (30, 2)
  (60, 3)
  (150, 8)
  (300, 16)
  (330, 5)
  (600, 30)
  (660, 11)
  (1155, 13)
  (1200, 61)
  (1650, 28)
  (2310, 26)
  (2805, 23)
  (3300, 66)
  (5610, 48)
  (5775, 71)
  (6600, 129)
  (10725, 77)
  (11550, 145)
  (13200, 261)
  (14025, 143)
  (21450, 155)
  (23100, 296)
  (28050, 268)
  (41925, 273)
  (46200, 673)
  (53625, 381)
  (56100, 586)
  (83850, 575)
  (107250, 853)
  (112200, 1254)
  (209625, 1434)
  (214500, 1741)
  (419250, 3003)
  (429000, 3677)
  (838500, 5974)
};
   \addlegendentry{\mimir};
\end{loglogaxis}
\end{tikzpicture}
    %\Description{Gaussian Mixture Model results}
    \vspace{-2.5ex}
    \caption{Runtime of the Gaussian Mixture Model (GMM) algorithm on different input sizes (lower is better)}
%The diagram is a log-log plot.
    \label{fig:gmm}
    \vspace{-3ex}
\end{figure}

\torchscript improves upon \pytorch by compiling parts of the code.
The resulting speedup is especially visible for very large input sizes where \torchscript{} becomes one of the best implementations as shown in \autoref{fig:gmm}.
\pytorch{} 2.0 further refines the \torchscript approach of ahead-of-time compilation using \toolname{TorchDynamo}~\cite{Ansel_TorchDynamo_2022}, \toolname{TorchInductor}, and \toolname{AOTAutoGrad}.
For small inputs, the remaining overhead of Python, not compiled functions, and communication overhead still causes \pytorch{} 2.0 to be slower than \enzyme{} and \thornado.
One advantage of \pytorch{} is its support for utilizing multiple cores.
In \autoref{fig:gmm}, the \pytorch{} 2.0 Multi-Core variant shows how this notably helps \pytorch's performance on large problem sizes.
%Only the \pytorch{} 2.0 benchmark utilizes multiple CPU cores and therefore performs better for larger problem sizes---using a single thread \pytorch{} 2.0 slows down notably.
%This is observable in the slowdown when forcing \pytorch{} 2.0 to use a single thread.
\enzyme{} and \thornado{} both do not use multiple cores yet.

\subsubsection{Code Complexity}
\label{sec:eval_complexity}

% LoC
% Hallstead Metric

%\begin{table}
%    \centering
%    \begin{tabular}{|c|r|r|}
%        \hline
%        Project & LoC Header & LoC Source \\
%        \hline
%        \hline
%        \enzyme & $18600$ & $37700$
%         \\ \hline
%         MimIR AD & $700$ & $100$
%         \\ \hline
%    \end{tabular}
%    \caption{Lines of code rounded to the next hundred (lower is better)}
%    \label{tab:loc}
%\end{table}
%
%
%\begin{table}
%    \centering
%\begin{tabular}{|c|r|r|r|}
%    \hline
%    Project & total & maximum & average \\
%    \hline
%    \hline
%    \enzyme & $13512$ & $892$ & $13.09$
%     \\ \hline
%     MimIR AD & $76$ & $13$ & $1.58$
%     \\ \hline
%\end{tabular}
%    \caption{Cyclomatic complexity metric of both implementations (lower is better)}
%    \label{tab:cyclomatic}
%\end{table}

One major contribution of our approach to \ac{AD} is its simplicity.
In order to verify this claim, we estimate the code complexity of \thornado{} and \enzyme{} using code complexity metrics.
\autoref{tab:loc_cyclomatic} summarizes the \ac{LoC} and the \emph{cyclomatic complexity}~\cite{mccabe1976complexity}.
For \enzyme we measured the source folder.
For better comparability, we also measured the metrics for \enzyme without the Clang plugin, the \mlir code, and without the different scalar evolution expander versions.
The numbers without extensions refer to the plugin without the special casing of pointer arguments.
\enzyme has roughly $10\times$ more code than \thornado.
This does not necessarily give an indication of which code is simpler.
But as a rule of thumb the larger a code base gets, the harder it becomes to debug and maintain.
As a more profound code complexity metric, we also look at the \emph{cyclomatic complexity} metrics\footnote{We used \href{https://github.com/metrixplusplus/metrixplusplus}{metrix++} for the measurement.} of both implementations.
\enzyme's cyclomatic complexity is roughly an order of magnitude larger than \thornado's.

\subsection{Other Plugins, Composition of Plugins, and Practical Experience}\label{sec:eval:discussion}

In addition, we have also created the following plugins:
\texttt{\%affine} to represent affine loop nests,
\texttt{\%clos} which implements a typed closure conversion~\cite{DBLP:conf/popl/MinamideMH96},
\texttt{\%direct} which allows to invoke direct-style functions as continuations and vice versa,
\texttt{\%refly} which allows for introspection \& reflection,
\texttt{\%compile} exposes \mimir's optimization framework as axioms, and
\texttt{\%opt} which implements \mimir's standard optimization pipeline as a \mim program.

We have already described the interaction of the \texttt{\%autodiff} and the \texttt{\%tensor} plugin in \autoref{sec:autodiff}.
In addition, the \texttt{\%autodiff} plugin uses the \texttt{\%affine} plugin for loops, the \texttt{\%clos} plugin to introduce and eliminate closures, and the \texttt{\%direct} plugin to seamlessly switch between direct style and \ac{CPS}.
The \texttt{\%core}, \texttt{\%math}, and \texttt{\%mem} plugins use each other mutually and are used by all other plugins that need to generate low-level code including the \texttt{\%regex}, \texttt{\%tensor}, and \texttt{\%autodiff} plugins.

During development, \mimir's type system was instrumental in identifying and addressing typical bugs such as incorrect wiring of expressions early in the implementation of our plugins (see \autoref{sec:cpp}).
In addition, MimIR's support for polymorphism enabled a shift from C++ code generating \mimir to directly working with polymorphic, type-safe \mim (see \autoref{sec:loop-unroll}).
For example, many plugins implement small wrappers and glue code directly in \mim.

% vim:spell:spelllang=en
\section{Related \& Future Work}
\label{sec:relwork}

\mimir lies at the intersection of higher-order, dependent type theory and low-level compiler \acp{IR} and is to the best of our knowledge the first of its kind in this regard.
We have already discussed \mimir's relationship to \thorin and \mlir in \autoref{sec:mimir_vs_thorin}.
This section discusses work that influenced \mimir and other related approaches.

\paragraph{Partial Evaluation \& AnyDSL}

Partial Evaluation filters were first introduced by SCHISM~\cite{DBLP:conf/esop/Consel88} and are also used by the modern partial evaluation framework AnyDSL~\cite{DBLP:journals/pacmpl/LeissaBHPMSMS18}.
AnyDSL has been successfully applied for high-performance applications such as sequence alignment~\cite{DBLP:conf/ipps/MullerS0MLKH20,DBLP:conf/ics/MullerSMLH22} or ray tracing~\cite{DBLP:journals/tog/Perard-GayotMLH19}.
AnyDSL's \ac{IR} is \thorin~\cite{DBLP:conf/cgo/LeissaKH15} (see \autoref{sec:overview:discussion}).
AnyDSL relies on \emph{shallow} \ac{DSL} embedding and partial evaluation to remove the overhead \ac{DSL} interfaces impose.
Like AnyDSL, \mimir supports shallow \ac{DSL} embedding but adds the possibility of \emph{deep embeddings} via plugins (\autoref{sec:overview}).
%\mimir~1 also lacks \mimir~2's optimizer that makes partial evaluation much more aggressive than in \mimir~1.
Furthermore, \mimir is to the best of our knowledge the first system that employs filter-based partial evaluation to resolve $\beta$-equivalence during type-checking a dependently typed language.

\paragraph{Sea of Nodes}

While the idea of a \enquote{sea of nodes} goes back to \citet{DBLP:conf/irep/ClickP95}, \thorin pioneered this concept for higher-order languages.
\mimir inherits this representation and simplifies the graph even further:
Due to \mimir's roots in \ac{PTS}, types---which are also just expressions---are part of the normal program graph as well.

Since the arrival of graph-based \acp{IR}, a debate has arisen among compiler engineers as to whether graph-based \acp{IR} or instruction lists are superior.
Graph-based \acp{IR} are used, for example, in the HotSpot JVM~\cite{DBLP:conf/jvm/PalecznyVC01}, GraalVM~\cite{DBLP:conf/oopsla/DuboscqWSWSM13}, and Google's TurboFan compiler (part of V8).
Graph-based \acp{IR} enable several optimizations directly during \ac{IR} construction, such as
constant folding,
various arithmetic simplifications,
or semi-global value numbering\footnote{This is \emph{global} in the sense that instructions float beyond basic block boundaries but \emph{local} as hashing stops at $\phi$-functions.} through hash-consing.
With graph-based \acp{IR}, operations track dependencies solely through data dependencies.
This makes the graphs invariant to code motion and allows analyses to avoid dead code by only following data dependence edges.
\mimir leverages these features for its normalization framework.
Instruction lists, in contrast, are simpler to understand and straightforward to traverse.
Visualizing a graph-based \ac{IR} (especially if it is broken) requires external graph tools and specialized debugging infrastructure.
In contrast, dumping an instruction list is straightforward---even if the program is incomplete or contains errors.
When the original order of instructions is crucial (e.g., when handling side effects), graph-based \acp{IR} must make these dependencies explicit.
This can feel cumbersome, but it also more accurate (see \autoref{sec:overview_types} and \ref{sec:mem_plugin}).

\paragraph{$\lambda$-Calculi \& SSA}

As a $\lambda$-calculus, \mimir takes inspiration from and shares similarities with many other $\lambda$-calculi---most notably:
\ac{CC}, $\lambda$-cube~\cite{DBLP:journals/jfp/Barendregt91}, \ac{PTS}, Henk~\cite{meijer1997henk}, and the zip calculus~\cite{DBLP:conf/mpc/Tullsen00}.
Since \mimir is based upon a predicative flavor of \ac{CC} it also subsumes a predicative flavor of System~F~\cite{DBLP:conf/programm/Reynolds74,GIRARD197163} and System~F$_\omega$.
\citet{DBLP:journals/jfp/RossbergRD14} have shown that ML's module system can be encoded in System~F$_\omega$.
Apart from impredicative idioms, \mimir can thus encode ML modules as well.

\mimir is also flexible enough to mimic various \ac{SSA} flavors.
We have already discussed in \autoref{sec:rec} how \ac{CPS} makes \mimir akin to an \ac{SSA} representation~\cite{DBLP:journals/sigplan/Appel88,DBLP:conf/irep/Kelsey95}.
\mimir can also model various extensions to \ac{SSA} form such as \emph{(thin) gated \ac{SSA}}~\cite{DBLP:conf/lcpc/Havlak93,DBLP:conf/pldi/TuP95}, \ac{LCSSA}, or \ac{SSI} form~\cite{ananian99}.
It just depends on where additional variables are placed.

\citet{DBLP:conf/pldi/MaziarzELFJ21} present an algorithm to hash expressions modulo $\alpha$-equivalence.
This technique does not work for \mimir, however, as \mimir's program graph is mutable at very specific spots (\autoref{sec:graph}).
Moreover, the \emph{assignable} relation checks for compatible tuple types (\autoref{sec:assignable}) and resolves implicits (\autoref{sec:infer}) anyway.
By doing this, the relation additionally checks for $\alpha$-equivalence.

\paragraph{CPS vs. Direct Style}

In the community for compilers of functional languages, there is a decades-old debate about whether to use
\ac{CPS}~\cite{DBLP:conf/sigplan/KranzKRHP86,DBLP:books/cu/Appel1992,DBLP:conf/icfp/Kennedy07} or
direct style~\cite{DBLP:conf/pldi/FlanaganSDF93,DBLP:conf/pldi/MaurerDAJ17}.
On the one hand, many optimizations such as simple rewrites like $x + 0 \vtr x$ are much easier to implement in direct-style.
On the other hand, we need \ac{CPS} to model unstructured control flow and at the end of the day, a compiled program consists of basic blocks with machine instructions that jump to each other---which \emph{is} \ac{CPS}.
Thus, we second the opinion of \citet{DBLP:journals/pacmpl/CongOER19} that the question of whether to compile with or without continuations is more a matter of what a compiler engineer/language designer wants to achieve and the specific stage in the compilation pipeline.
\mimir itself does not prioritize the use of \ac{CPS} or direct style, offering syntactic sugar to accommodate both styles.
The \mimir plugin \texttt{\%direct} enables the invocation of direct-style functions as continuations and certain continuations as direct-style functions.
For example, given \lstinline|f| of type \lstinline|Fn T -> U|, the expression \lstinline|%direct.cps2ds f| has type \lstinline|T -> U|.
Here we noted similarities to negation in the work of \citet{DBLP:journals/pacmpl/OstermannBSSD22}.
In addition, the \texttt{\%direct} plugin \ac{CPS}-converts direct-style function to \ac{CPS} because \mimir's \llvm backend expects \ac{CPS}.
\citet{DBLP:journals/pacmpl/CongOER19} present a calculus that differentiates between first- and second-class continuations and a type system that ensures proper use.
\mimir's \texttt{\%clos} plugin performs a similar classification via static analysis, instead.
This plugin transforms escaping continuations via typed closure conversion~\cite{DBLP:conf/popl/MinamideMH96} but is limited to non-dependent function types.
\citet{DBLP:conf/pldi/BowmanA18} present a technique to closure-convert dependently typed functions in \ac{CC}.

\paragraph{Typed Assembly Language}

Previous work on low-level types~\cite{DBLP:conf/popl/MorrisettWCG98} enriched imperative assembly languages with a type system.
Building up on this work, \ac{DTAL} adds dependent types that are limited to linear constraints on integers while \mimir features full dependent types.
\ac{DTAL} uses an \ac{ILP} solver to solve these constraints whereas \mimir uses normalization and partial evaluation filters to decide dependent type checking.
These strategies are orthogonal to each other and in \mimir we could define a \texttt{\%dtal} plugin that captures \ac{DTAL}-like constraints in an axiomatized dependent type and connect to an \ac{ILP} solver during normalization to find more program equivalences and, hence, emulate \ac{DTAL} constraints.

\paragraph{Dependent Types}
Dependent types give more freedom in program construction and allow to express semantic properties in the types of expressions.
Although they have been mainly used in proof assistants such as Coq, Agda, or Lean in the past, they have received some attention beyond proof assistants in recent years: % TODO citations?
Dependent Haskell~\cite{DBLP:journals/pacmpl/WeirichVAE17} is an extension of Haskell that allows combining types and expressions while preserving backward compatibility with Haskell.
Scala~3 is based upon \ac{DOT}~\cite{DBLP:conf/birthday/AminGORS16} which features \emph{path-dependent types}.
These are a specific kind of dependent type where the dependent-upon value is a path.

Idris focuses on type-driven development but, if desired, properties of program behavior can be formally stated and proven.
\mimir similarly expresses semantics on the type-level using dependent-types but differs in the focus on optimization instead of verification. Whereas Idris is a high-level programming language, \mimir as an \ac{IR} is much closer to the hardware.
Idris~2~\cite{DBLP:conf/ecoop/Brady21} introduces \ac{QTT}.
This not only allows to specify protocols as types but also to tag which types should be erased to clearly state which parts of a program materializes at runtime.
\mimir on the other hand, uses partial evaluation filters to resolve $\beta$-equivalence and make this distinction.

F*~\cite{DBLP:conf/popl/SwamyHKRDFBFSKZ16} combines automated theorem proving and interactive proving.
The automation also aids when using refinement types as utilized in length-annotated arrays.
\mimir's user experience is similar as the normalization approach handles dependent types without further complications.
However, \mimir does not aim to be a theorem prover. Furthermore, \mimir uses general dependent types instead of refinement types. This approach allows for more freedom as is exemplified in the \ac{AD} plugin that computes the result type at the type level.
Furthermore, the developer can extend the normalization approach to handle more complex cases.
One possible rule could optimize a double reversion \lstinline|rev (rev xs)| directly to \lstinline|xs|; a use case, many automatic systems have problems with.

\paragraph{Future Work}

In the future, we want to make it possible to specify normalizations directly as rewrite rules in \mim.
This further reduces boilerplate C++ code.
Moreover, this opens the door for additional sanity checks and nondeterministic rule sets that we could explore with rewrite engines like equality saturation~\cite{DBLP:conf/popl/TateSTL09}.
For example, map/reduce-style rewrite rules have already been used with equality saturation to produce high-performance code~\cite{DBLP:journals/corr/abs-2111-13040}.
We are also working on support for singleton, union, and intersection types.
Singleton and union types are useful in many settings, e.g., we could replace the type of the \lstinline|mode| arguments in the \verb|%core| and \verb|%math| dialect with a proper union type instead of a \lstinline|Nat|.
Intersection types allow us to combine trait-like $\Sigma$-types such as \lstinline|Cmp| $\cap$ \lstinline|Num|.
We also want to add linear types or full \ac{QTT} in order to validate by the type checker whether values that are supposed to be used linearly are used correctly.
For example, values of type \lstinline|%mem.M|---which track side-effects---must be used linearly, but this is right now not enforced by the type system.
Furthermore, we want to enable parallelization and accelerators, such as GPUs, to enhance higher-level \acp{DSL} with these mechanisms.
Finally, we want to reimplement existing \acp{DSL} such as Lift~\cite{DBLP:conf/cgo/SteuwerRD17} or RISE/Shine~\cite{DBLP:journals/corr/abs-2201-03611} in \mimir.

% vim:spell:spelllang=en
\section{Conclusion}
\label{sec:concl}

In this paper, we presented \mimir, an extensible, higher-order intermediate representation.
\mimir's extensibility allows for expressing and optimizing programs at any level of abstraction.
\mimir's foundation in \acl{CC} provides a general and common type system for domain-specific languages to nest in.
\ac{DSL} authors benefit from reusing \mimir's type system, normalization, and optimization framework without having to provide manual type-checkers and reimplementing standard optimizations for their custom operations.
We have shown that using \mimir, we can generate code with state-of-the-art performance for high-level, domain-specific applications such as a \ac{RegEx} matcher, automatic differentiation, as well as for low-level, imperative code.

\section*{Data Availability}

\mimir is available as open source on GitHub.\footnote{see \url{https://anydsl.github.io/MimIR/}}
All data used in \autoref{sec:eval} is available at Zenodo~\cite{leissa_2024_13952579}.

\begin{acks}
    The authors would like to thank the anonymous reviewers for their concise and constructive feedback, which has greatly contributed to the improvement of this paper.
\end{acks}

\bibliographystyle{ACM-Reference-Format}
\bibliography{dblp,other,autodiff}

\end{document}